\documentclass[useAMS,usenatbib,usegraphicx,fleq]{mn2e} 
\usepackage{times}
\usepackage{natbib}
\usepackage{amssymb,amsmath,bm}
\usepackage[colorlinks=true]{hyperref}
\usepackage{graphicx}
\usepackage{caption}
\usepackage{subcaption}
\usepackage{epstopdf} 
\usepackage{multirow}
\graphicspath{{./}}
\usepackage{color}
\usepackage[plain]{fancyref}
\usepackage[nolist]{acronym}
\usepackage{soul} %

\makeatletter
\newcommand{\extraclearlabels}
{\protected@write\@auxout{}{%
\string\reset@newl@bel
}}
\makeatother

\newcommand{\Munich}{$^{1}$}
\newcommand{\ExcellenceCluster}{$^{2}$}
\newcommand{\MPE}{$^{3}$}
\newcommand{\UChicago}{$^{4}$}
\newcommand{\CfA}{$^{5}$}
\newcommand{\MIT}{$^{6}$}
\newcommand{\Harvard}{$^{7}$}
\newcommand{\FNAL}{$^{8}$}
\newcommand{\KICPChicago}{$^{9}$}
\newcommand{\AAUChicago}{$^{10}$}
\newcommand{\PhysicsUChicago}{$^{11}$}
\newcommand{\ANL}{$^{12}$}
\newcommand{\Miss}{$^{13}$}
\newcommand{\EFIChicago}{$^{14}$}
\newcommand{\NIST}{$^{15}$}
\newcommand{\PUC}{$^{16}$}
\newcommand{\Caltech}{$^{17}$}
\newcommand{\McGill}{$^{18}$}
\newcommand{\illast}{$^{19}$}
\newcommand{\illphy}{$^{20}$}
\newcommand{\Berkeley}{$^{21}$}
\newcommand{\UFlorida}{$^{22}$}
\newcommand{\Colorado}{$^{23}$}
\newcommand{\KavliCA}{$^{24}$}
\newcommand{\Stanford}{$^{25}$}
\newcommand{\LBNL}{$^{26}$}
\newcommand{\Michigan}{$^{27}$}
\newcommand{\Minnesota}{$^{28}$}
\newcommand{\Melbourne}{$^{29}$}
\newcommand{\STScI}{$^{30}$}
\newcommand{\CaseWestern}{$^{31}$}
\newcommand{\SAIC}{$^{32}$}
\newcommand{\Dunlap}{$^{33}$}
\newcommand{\Toronto}{$^{34}$}
\newcommand{\BCCP}{$^{35}$}
\newcommand{\CTIO}{$^{36}$}

\newcommand{\sdeg} {\text{deg}^{2}}

\newcommand{\lx}{L_{\text X}}
\newcommand{\SN}{\xi_{\text X}}
\newcommand{\asz}{A_{\text{SZ}}}
\newcommand{\bsz}{B_{\text{SZ}}}
\newcommand{\csz}{C_{\text{SZ}}}
\newcommand{\dsz}{D_{\text{SZ}}}
\newcommand{\planck}{{\it Planck}}
\newcommand{\xbcs}{XMM-BCS}
\newcommand{\plx}{P11}
\newcommand{\ysz}{Y_\text{500}}
\newcommand{\rfive}{R_\text{500c}}
\newcommand{\rc}{R_\text{c}}
\newcommand{\ay}{A_\text{Y}}%
\newcommand{\by}{B_\text{Y}} %
\newcommand{\msun}{\mathrm{M_{\odot}}} %
\newcommand{\avg}[1]{\langle #1 \rangle}
\newcommand{\mr}[1]{\mathrm{#1}}
\newcommand{\rsz}{\bm{r}_\text{SZ}}
\newcommand{\rx}{\bm{r}_\text{X}}
\newcommand{\tx}{\Theta_\text{X}}
\begin{document}

\title[SZE-Mass Relations in Low Mass Clusters and Groups]
{Analysis of Sunyaev-Zel'dovich Effect Mass-Observable Relations using South Pole Telescope Observations of an X-ray Selected Sample of Low Mass Galaxy Clusters and Groups}

\author[J. Liu et al.] {J.~Liu\Munich$^,$\ExcellenceCluster,
J.~Mohr\Munich$^,$\ExcellenceCluster$^,$\MPE,
A.~Saro\Munich$^,$\ExcellenceCluster, 
K.~A.~Aird\UChicago,
M.~L.~N.~Ashby\CfA,
M.~Bautz\MIT,
M.~Bayliss\Harvard$^,$\CfA,
\newauthor 
B.~A.~Benson\FNAL$^,$\KICPChicago$^,$\AAUChicago,
L.~E.~Bleem\KICPChicago$^,$\PhysicsUChicago$^,$\ANL,
S.~Bocquet\Munich$^,$\ExcellenceCluster,
M.~Brodwin\Miss,
J.~E.~Carlstrom\KICPChicago$^,$\AAUChicago$^,$\PhysicsUChicago$^,$\ANL$^,$\EFIChicago,
\newauthor 
C.~L.~Chang\KICPChicago$^,$\EFIChicago$^,$\ANL,
I.~Chiu\Munich$^,$\ExcellenceCluster,
H.~M. Cho\NIST,
A.~Clocchiatti\PUC,
T.~M.~Crawford\KICPChicago$^,$\AAUChicago,
\newauthor 
A.~T.~Crites\KICPChicago$^,$\AAUChicago$^,$\Caltech,
T.~de~Haan\McGill,
S.~Desai\Munich$^,$\ExcellenceCluster,
J.~P.~Dietrich\Munich$^,$\ExcellenceCluster,
M.~A.~Dobbs\McGill,
R.~J.~Foley\CfA$^,$\illast$^,$\illphy,
\newauthor 
D.~Gangkofner\Munich$^,$\ExcellenceCluster,
E.~M.~George\Berkeley,
M.~D.~Gladders\KICPChicago$^,$\AAUChicago,
A.~H.~Gonzalez\UFlorida,
N.~W.~Halverson\Colorado,
\newauthor
C.~Hennig\Munich$^,$\ExcellenceCluster,
J.~Hlavacek-Larrondo\KavliCA$^,$\Stanford,
G.~P.~Holder\McGill,
W.~L.~Holzapfel\Berkeley,
J.~D.~Hrubes\UChicago,
\newauthor
C.~Jones\CfA,
R.~Keisler\KICPChicago$^,$\PhysicsUChicago,
A.~T.~Lee\Berkeley$^,$\LBNL,
E.~M.~Leitch\KICPChicago$^,$\AAUChicago,
M.~Lueker\Berkeley$^,$\Caltech,
D.~Luong-Van\UChicago,
\newauthor
M.~McDonald\MIT,
J.~J.~McMahon\Michigan,
S.~S.~Meyer\KICPChicago$^,$\EFIChicago$^,$\PhysicsUChicago$^,$\AAUChicago,
L.~Mocanu\KICPChicago$^,$\AAUChicago,
S.~S.~Murray\CfA,
\newauthor 
S.~Padin\KICPChicago$^,$\AAUChicago$^,$\Caltech,
C.~Pryke\Minnesota,
C.~L.~Reichardt\Berkeley$^,$\Melbourne
A.~Rest\STScI,
J.~Ruel\Harvard,
J.~E.~Ruhl\CaseWestern,
\newauthor
B.~R.~Saliwanchik\CaseWestern,
J.~T.~Sayre\CaseWestern,
K.~K.~Schaffer\KICPChicago$^,$\EFIChicago$^,$\SAIC,
E.~Shirokoff\Berkeley$^,$\Caltech,
H.~G.~Spieler\LBNL,
\newauthor
B.~Stalder\CfA,
Z.~Staniszewski\CaseWestern$^,$\Caltech,
A.~A.~Stark\CfA,
K.~Story\KICPChicago$^,$\PhysicsUChicago,
R.~\v Suhada\Munich,
K.~Vanderlinde\Dunlap$^,$\Toronto,
\newauthor
J.~D.~Vieira\illast$^,$\illphy,
A. Vikhlinin\CfA,
R.~Williamson\KICPChicago$^,$\AAUChicago$^,$\Caltech,
O.~Zahn\Berkeley$^,$\BCCP,
A.~Zenteno\Munich$^,$\CTIO\\
\noindent
\Munich Department of Physics, Ludwig-Maximilians-Universit\"{a}t, Scheinerstr.\ 1, 81679 M\"{u}nchen, Germany \\
\ExcellenceCluster Excellence Cluster Universe, Boltzmannstr.\ 2, 85748 Garching, Germany \\
\MPE Max-Planck-Institut f\"{u}r extraterrestrische Physik, Giessenbachstr.\ 85748 Garching, Germany \\
\UChicago University of Chicago, 5640 South Ellis Avenue, Chicago, IL 60637 \\
\CfA Harvard-Smithsonian Center for Astrophysics, 60 Garden Street, Cambridge, MA 02138 \\
\Harvard Department of Physics, Harvard University, 17 Oxford Street, Cambridge, MA 02138 \\
\MIT Kavli Institute for Astrophysics and Space Research, Massachusetts Institute of Technology, 77 Massachusetts Avenue, Cambridge, MA 02139 \\
\FNAL Center for Particle Astrophysics, Fermi National Accelerator Laboratory, Batavia, IL, USA 60510 \\
\KICPChicago Kavli Institute for Cosmological Physics, University of Chicago, 5640 South Ellis Avenue, Chicago, IL 60637 \\
\AAUChicago Department of Astronomy and Astrophysics, University of Chicago, 5640 South Ellis Avenue, Chicago, IL 60637\\
\PhysicsUChicago Department of Physics, University of Chicago, 5640 South Ellis Avenue, Chicago, IL 60637 \\
\ANL Argonne National Laboratory, 9700 S. Cass Avenue, Argonne, IL, USA 60439 \\
\Miss Department of Physics and Astronomy, University of Missouri, 5110 Rockhill Road, Kansas City, MO 64110 \\
\EFIChicago Enrico Fermi Institute, University of Chicago, 5640 South Ellis Avenue, Chicago, IL 60637 \\
\NIST NIST Quantum Devices Group, 325 Broadway Mailcode 817.03, Boulder, CO, USA 80305 \\
\PUC Departamento de Astronomia y Astrosifica, Pontificia Universidad Catolica,Chile \\
\Caltech California Institute of Technology, 1200 E. California Blvd., Pasadena, CA 91125 \\
\McGill Department of Physics,McGill University, 3600 Rue University, Montreal, Quebec H3A 2T8, Canada \\
\illast Astronomy Department, University of Illinois at Urbana-Champaign,1002 W.\ Green Street,Urbana, IL 61801 USA \\
\illphy Department of Physics, University of Illinois Urbana-Champaign,1110 W.\ Green Street,Urbana, IL 61801 USA \\
\Berkeley Department of Physics, University of California, Berkeley, CA 94720 \\
\UFlorida Department of Astronomy, University of Florida, Gainesville, FL 32611 \\
\Colorado Department of Astrophysical and Planetary Sciences and Department of Physics, University of Colorado,Boulder, CO 80309 \\
\KavliCA  Kavli Institute for Particle Astrophysics and Cosmology, Stanford University, 452 Lomita Mall, Stanford, CA 94305-4085, USA \\
\Stanford Department of Physics, Stanford University, 452 Lomita Mall, Stanford, CA 94305-4085, USA \\
\LBNL Physics Division, Lawrence Berkeley National Laboratory, Berkeley, CA 94720 \\
\Michigan Department of Physics, University of Michigan, 450 Church Street, Ann Arbor, MI, 48109 \\
\Minnesota Physics Department, University of Minnesota, 116 Church Street S.E., Minneapolis, MN 55455 \\
\Melbourne School of Physics, University of Melbourne, Parkville, VIC 3010, Australia\\
\STScI Space Telescope Science Institute, 3700 San Martin Dr., Baltimore, MD 21218 \\
\CaseWestern Physics Department, Center for Education and Research in Cosmology and Astrophysics, Case Western Reserve University, Cleveland, OH 44106 \\
\SAIC Liberal Arts Department, School of the Art Institute of Chicago, 112 S Michigan Ave, Chicago, IL 60603 \\
\Dunlap Dunlap Institute for Astronomy \& Astrophysics, University of Toronto, 50 St George St, Toronto, ON, M5S 3H4, Canada \\
\Toronto Department of Astronomy \& Astrophysics, University of Toronto, 50 St George St, Toronto, ON, M5S 3H4, Canada\\
\BCCP Berkeley Center for Cosmological Physics, Department of Physics, University of California, and Lawrence Berkeley National Labs, Berkeley, CA 94720\\
\CTIO  Cerro Tololo Inter-American Observatory, Casilla 603, La Serena, Chile
}

\maketitle

\begin{acronym}
  \acrodef{sze}[SZE]{Sunyaev-Zel'dovich effect}
  \acrodef{cmb}[CMB]{cosmic microwave background}
  \acrodef{spt}[SPT]{South Pole Telescope}
  \acrodef{sz}[SZ]{Sunyaev-Zel'dovich} 
  \acrodef{xbcs}[XMM-BCS]{ {\it XMM-Newton} Blanco Cosmology Survey}
  \acrodef{sptsz}[SPT-SZ]{South Pole Telescope Sunyaev-Zel'dovich
    survey} 
  \acrodef{act}[ACT]{Atacama Cosmology Telescope}
  \acrodef{agn}[AGN]{active galactic nucleus}
  \acrodef{bcs}[BCS]{Blanco Cosmology Survey}
  \acrodef{mcmc}[MCMC]{Monte Carlo Markov Chain}
  \acrodef{sumss}[SUMSS]{Sydney University Molonglo Sky Survey}
  \acrodef{fwhm}[FWHM]{full-width half maximum}
  \acrodef{bcg}[BCG]{brightest cluster galaxy}
\end{acronym}
\begin{abstract}
\extraclearlabels 
We use microwave observations from the \ac{spt} to examine the \ac{sze} signatures of a sample of 46 X-ray selected groups and clusters drawn from $\sim6~\sdeg$ of the \ac{xbcs}.  These systems extend to redshift $z=1.02$ and probe the \ac{sze} signal to the lowest X-ray luminosities ($\ge$10$^{42}$~erg~s$^{-1}$) yet; these sample characteristics make this analysis complementary to previous studies.
We develop an analysis tool, using X-ray luminosity as a mass proxy, to extract selection-bias corrected constraints on the \ac{sze} significance- and $\ysz$-mass relations. The former is in good agreement with an extrapolation of the relation obtained from high mass clusters.  However, the latter, at low masses, while in good agreement with the extrapolation from the high mass \ac{spt} clusters, is in tension at 2.8\,$\sigma$ with the \planck\ constraints, indicating the low mass systems exhibit lower \ac{sze} signatures in the SPT data.  
We also present an analysis of potential sources of contamination.  For the radio galaxy point source population, we find 18 of our systems have 843~MHz \ac{sumss} sources within 2~arcmin of the X-ray centre, and three of these are also detected at significance $>$4 by \ac{spt}.   Of these three, two are associated with the group brightest cluster galaxies (BCGs), and the third is likely an unassociated quasar candidate.  We examine the impact of these point sources on our \ac{sze} scaling relation analyses and find no evidence of biases.  We also examine the impact of dusty galaxies using constraints from the 220~GHz data.  The stacked sample provides 2.8\,$\sigma$ significant evidence of dusty galaxy flux, which would correspond to an average underestimate of the \ac{spt} $\ysz$ signal that is $(17\pm9)$~per cent in this sample of low mass systems.  Finally, we explore the impact of future data from SPTpol and XMM-XXL, showing that it will lead to a factor of four to five tighter constraints on these \ac{sze} mass-observable relations.

\end{abstract}


\acresetall

\section{Introduction}

The \acl{sze} \cite[\acsu{sze}]{sunyaev70,sunyaev72}, is a spectral distortion of the \ac{cmb} arising from interactions between CMB photons and hot, ionised gas.  Surveys of galaxy clusters using the \ac{sze} have opened a new window on the Universe by providing samples of hundreds of massive galaxy clusters with well-understood selection over a broad redshift range.  Both space- and ground-based instruments, including the \planck\ satellite \citep{tauber10}, the \acl{spt} \acused{spt}\citep[\acs{spt};][]{carlstrom11}, and the Atacama Cosmology Telescope \citep[ACT;][]{fowler07}, have released catalogs of their \ac{sze} selected clusters.  The cluster samples have provided new cosmological constraints \cite[]{reichardt13, hasselfield13,planck13-20} and have enabled important evolution studies of cluster galaxies and the intracluster medium over a broad range of redshift \cite[e.g.,][]{zenteno11, semler12, mcdonald13}.

Understanding the relationship between the \ac{sze} observable and cluster mass is important for both cosmological applications and astrophysical studies. Among observables, the integrated Comptonization from the \ac{sze} has been shown by numerical simulations \cite[]{motl05,nagai07} to be a good mass proxy with low intrinsic scatter.  Cluster mass estimates derived from X-ray observations of \ac{sze} selected clusters have largely confirmed this expectation \cite[]{andersson11,planckearlyXI}. A related quantity, the \ac{spt} signal-to-noise $\xi$, is linked to the underlying virial mass of the cluster by a power law with log-normal scatter at the $\sim 20$~per cent level \cite[][hereafter B13]{benson13}.

Probing the \ac{sze} signature of low mass clusters and groups is also important, although it is much more challenging with the current generation of experiments. These low mass clusters and groups are far more numerous and are presumably important environments for the transformation of galaxies from the field to the cluster. Studies of their baryonic content show that low mass clusters and groups are not simply scaled-down versions of the more massive clusters \cite[e.g.,][]{mohr99, sun09, lagana13}. This breaking of self-similarity in moving from the cluster to the group mass scale is likely due to processes such as star formation and \ac{agn} feedback.

The \planck\ team has recently studied this low mass population by stacking the \planck\ maps around samples of X-ray selected clusters in the nearby universe \cite[][hereafter \plx]{planck11-10}. They show that the \ac{sze} signal is consistent with the self-similar scaling relation based on the X-ray luminosity over a mass range spanning 1.4 orders of magnitude.  

Here we pursue a study of the \ac{sze} signatures of low mass clusters extending over a broad range of redshift.  We use the \ac{sptsz} data with the \ac{xbcs} over $6~\sdeg$ from which a sample of 46 X-ray groups and clusters have been selected \cite[][hereafter S12]{suhada12}. The \ac{sptsz} data enable us to extract cluster \ac{sze} signal with high angular resolution and low instrument noise, making the most of this small sample.

The paper is organised as follows. In \Fref{sec:xbcs-data-description}, we describe the data used from the \ac{xbcs} and the extraction of the \ac{sze} signature from the \ac{sptsz} maps. In \Fref{sec:xbcs-method}, we introduce the calibration method for the mass-observable scaling relation, and we apply it to the cluster sample in \Fref{sec:xbcs-result}. We also discuss possible systematic effects and present a discussion of the point source population associated with our sample.  We conclude in \Fref{sec:xbcs-conclusions} with a prediction of the improvement based on future surveys.

The cosmological model parameters adopted in this paper are the same as the ones used for the X-ray measurement from the \ac{xbcs} project (S12): $(\Omega_{\mathrm M}, \Omega_{\Lambda},H_{0})=(0.3,0.7,70~\text{km s}^{-1}\text{Mpc}^{-1})$. The amplitude of the matter power spectrum, which is needed to estimate bias corrections in the analysis, is fixed to $\sigma_{8}=0.8$.

\section{Data Description and Observables}\label{sec:xbcs-data-description}

In this analysis, we adopt an X-ray selected sample of clusters, described in \Fref{sec:xbcs-xmmbcs}, together with published $\lx$-mass scaling relations to examine the corresponding \acsu{sptsz} significance- and $\ysz$-mass relations. The \ac{sptsz} observable $\xi$ is measured by a matched filter approach, which we discuss in \Fref{sec:xbcs-spt} and \Fref{sec:xbcs-spt-xi}.  The estimation of $\ysz$ is described in \Fref{sec:xbcs-yintegrated}.

\subsection{X-ray Catalog}\label{sec:xbcs-xmmbcs}

The \ac{xbcs} project consists of an X-ray survey mapping $14~\sdeg$ area of the southern hemisphere sky that overlaps the {\it griz} bands \acl{bcs} \cite[\acsu{bcs}]{desai12} and the mm-wavelength \ac{sptsz} survey \cite[]{carlstrom11}. S12 analyse the initial $6~\sdeg$ core area, construct a catalog of 46 galaxy clusters and present a simple selection function. Here we present a brief summary of the characteristics of that sample.  The cluster physical parameters from Table 2 (S12) are repeated in \Fref{tab:xbcs-cat} with the same IDs.

The initial cluster sample was selected via a source detection pipeline in the 0.5--2~keV band.  The spatial extent of the clusters leads to the need to have more counts to reach a certain detection threshold than are needed for point sources.  S12 modelled the extended source sensitivity as an offset from the point source limit; the cluster sample is approximately a flux-limited sample with $f_\text{min}=1\times10^{-14} \text{erg s}^{-1}\text{cm}^{-2}$.  

The X-ray luminosity $\lx$ was measured in the detection band (0.5\ -\ 2.0~keV) within a radius of $\rfive$, which is iteratively determined using mass estimates from the $\lx$-mass relation and is defined such that the interior density is 500~times the critical density of the Universe at the corresponding  redshift.  This luminosity was converted to a bolometric luminosity and to a 0.1\ -\ 2.4~keV band luminosity using the characteristic temperature for a cluster with this 0.5\ -\ 2.0~keV luminosity and redshift (see equation~3 in S12).  The core radius, $R_\text{c}$, of the beta model is calculated using (see equation~1 in S12):
\begin{equation}
R_\text{c} = 0.07\times R_{500} \Big(\frac{T}{\text{1~keV}}\Big)^{0.63},
\end{equation}
where $T$ is X-ray temperature determined through the $\lx-T$ relation.  
The redshifts of the sample are primarily photometric redshifts extracted using the \ac{bcs} optical imaging data. The optical data and their processing and calibration are described in detail elsewhere \cite[]{desai12}. The photometric redshift estimator has been demonstrated on clusters with spectroscopic redshifts and on simulations \cite[]{song12a} and has been used for redshift estimation within the \ac{sptsz} collaboration \cite[]{song12b}. The typical photometric redshift uncertainty in this \ac{xbcs} sample is $\avg{\Delta z/(1+z)}=0.023$, which is determined using a subsample of 12 clusters ($z<0.4$) with spectroscopic redshifts.  This value is consistent with the uncertainty $\avg{\Delta z/(1+z)}=0.017$ we obtained on the more massive main sample \ac{sptsz} clusters.

\begin{figure}
  \includegraphics[width=3.5in]{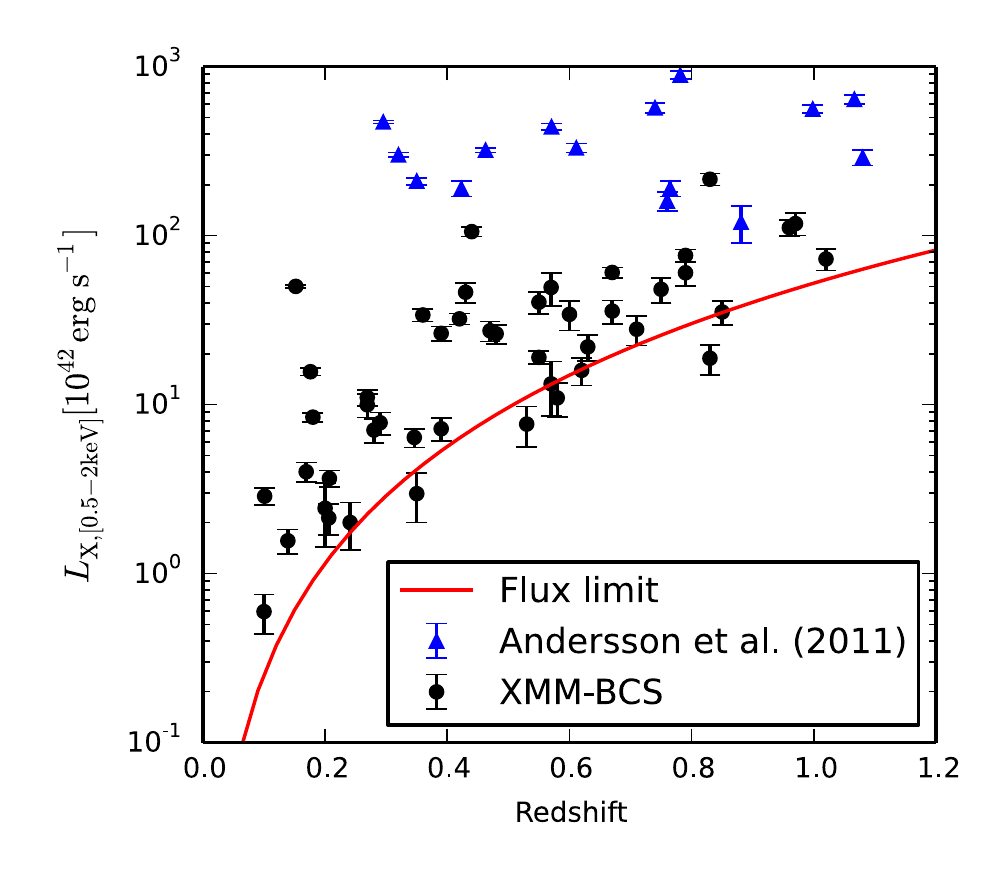}
  \caption[The luminosity-redshift distribution of the XMM-BCS
    clusters from \protect\citet{suhada12} and the \ac{sptsz} clusters from
    \protect\citet{andersson11}]{The luminosity-redshift distribution of the \ac{xbcs}
    clusters from S12 (black dots) and the \ac{sptsz} clusters from
    \protect\citet[blue triangles]{andersson11}.  The X-ray sample is
    selected with a flux cut that varies somewhat across the field.  
    The red line is the corresponding luminosity sensitivity determined by the median flux limit in the 0.5 - 2.0~keV band.  The \ac{sptsz} sample is more massive and approximately mass limited.}
  \label{fig:xbcs-lxza}
\end{figure}

The X-ray luminosities and photometric redshifts of the sample are shown in \Fref{fig:xbcs-lxza} in black squares and the approximate flux limit of the sample is shown as a red curve.  For comparison, we also include a  high mass \ac{sptsz} cluster sample (blue triangles) with published X-ray properties \cite[]{andersson11}.

In the analysis that follows we use the X-ray luminosity as the primary mass estimator for each cluster.  We adopt the $\lx$-mass scaling relation used in S12, which is based on the hydrostatic mass measurements in an ensemble of 31 nearby clusters observed with {\it XMM-Newton} \cite[{REXCESS},][]{pratt09}:
\begin{equation} \label{eq:xbcs-lm}
  \lx=L_{0}\Big(\frac{M_\text{500c}}{2\times10^{14}\msun}
  \Big)^{\alpha_\text{LM}}E(z)^{7/3},
\end{equation}
where $H(z)=H_0 E(z)$.   The intrinsic scatter in $\lx$ at fixed mass is modelled as lognormal distributions with widths $\sigma_{\lx}$, and the observational scatter is given in S12.

This scaling relation includes corrections for Malmquist and Eddington biases.  Both biases are affected by the intrinsic scatter and the skewness of the underlying sample distribution. In general, the bias on the true mass is $\Delta \ln M\ \propto\ \gamma \sigma_{\ln M}^2$, where $\mr{d}n(M)/\mr{d}\ln M \propto M^\gamma$ is the slope of the mass distribution and $\sigma_{\ln M}$ is the scatter in mass at fixed observable \citep[for more discussion, we refer the reader to][]{stanek06,vikhlinin09b,mortonson11}. Typically $\gamma$ is negative, and the result is that mass inferred from an observable must be corrected to a lower value than that suggested by naive application of the scaling relation.

The scaling relation parameters for different X-ray bands are listed in \Fref{tab:xbcs-lxm}.  We find the choice of luminosity bands has negligible impact on the parameter estimation given the current constraint precision.  In addition, we  investigate using the $\lx$-mass scaling relations from {\it  Chandra} observations \citep{vikhlinin09b,mantz10b}.  These studies draw upon higher mass
cluster samples than the {REXCESS} sample, and therefore we adopt the \citet{pratt09} relation for our primary analysis.  We discuss the impact of changing the $\lx$-mass scaling relation in \Fref{sec:xbcs-xim}.

\begin{table} \begin{center} \caption{$\lx$-mass relations with
      different luminosity bands
      (\Fref{eq:xbcs-lm}).} \label{tab:xbcs-lxm}
    \begin{tabular}[h]{|c|c|c|c|} \hline Type &
      $L_{0}[10^{44}\text{erg\ s}^{-1}]$& $\alpha_\text{LM}$ & $\sigma_{\ln \lx}$\\ \hline
      0.5--2.0 keV & $0.48\pm0.04$ & $1.83\pm0.14$ & $0.412\pm0.071$ \\
      0.1--2.4 keV & $0.78\pm0.07$ & $1.83\pm0.14$ & $0.414\pm0.071$ \\
      Bolometric & $1.38\pm0.12$ & $2.08\pm0.13$ &
      $0.383\pm0.061$ \\ \hline \end{tabular}
  \end{center} \end{table}

\subsection{SPT Observations} \label{sec:xbcs-spt}

The \ac{spt} \citep{carlstrom11} is a 10-metre diameter, millimetre-wavelength, wide field telescope that was deployed in 2007 and has been used since then to make arcminute-resolution observations of the CMB over large areas of the sky. The high angular resolution is crucial to detecting the \ac{sze} signal from high-redshift clusters. The \ac{sptsz} survey \citep[e.g.,][]{story13}, completed in 2011, covers a 2500 deg$^2$ region of contiguous sky area in three bands -- centred at 95, 150, and 220 GHz -- at a typical noise level of $< 18 \mu$K per one-arcminute pixel in the 150 GHz band.

The details of the \ac{sptsz} observation strategy, data processing and mapmaking are documented in \cite{schaffer11}; we  briefly summarise them here. The \ac{sptsz} survey data were taken primarily in a raster pattern with azimuth scans at discrete elevation steps. A high-pass filter was applied to the time-ordered data to remove low-frequency atmospheric and instrumental noise. The beams, or angular response functions, were measured using observations of planets and bright \acp{agn} in the field. The main lobe of the beam for a field observation is well-approximated as a Gaussian with a \ac{fwhm} of 1.6, 1.2, and 1.0~arcmin at 95, 150, and 220~GHz, respectively. The final temperature map was calibrated by the Galactic H\,{\sevensize\sc II} regions RCW38 and MAT5a \cite[c.f.][]{vanderlinde10}.  The \ac{sptsz} maps used in this work are from a 100 deg$^2$ field centred at $(\alpha, \delta)$ = $(23^\circ\ 30', -55^\circ)$ and consist of observations from the 2008 and 2010 \ac{sptsz} observing seasons.  The characteristic depths are 37, 12 and $35~\mu\text{K-arcmin}$ at 95, 150 and 220~GHz, respectively.

\subsection{SPT-SZ Cluster Significance}\label{sec:xbcs-spt-xi}

The process of determining the \ac{sptsz} significance for our X-ray sample is very similar to the process of finding clusters in \ac{sptsz} maps, but there are certain key differences, which we highlight below.  Clusters of galaxies are extracted from \ac{sptsz} maps through their distinct angular scale- and frequency-dependent imprint on the \ac{cmb}. We adopt the multi-frequency matched filter approach \cite[]{melin06} to extract the cluster signal. The matched filter is designed to maximise the given signal profile while suppressing all noise sources. A detailed description appears elsewhere \citep{vanderlinde10, williamson11}. Here we provide a summary.  The \ac{sze} introduces a spectral distortion of the \ac{cmb} at given
frequency $\nu$ as:
\begin{equation}
  \label{eq:xbcs-cmby}
  \Delta T_\text{CMB}(\bm{\theta},\nu)  = y(\bm{\theta}) g(\nu) T_\text{CMB},
\end{equation}
where $g(\nu)$ is the frequency dependency and the Compton-y parameter $y(\bm{\theta})$ is the \ac{sze} signature at direction $\bm{\theta}$, which is linearly related to the integrated pressure along the line-of-sight.  To model the \ac{sze} signal $y(\bm{\theta})$, two common templates are adopted: the circular $\beta$ model \cite[]{cavaliere76} and the Arnaud profile \cite[]{arnaud10}.  The cluster profiles are convolved with the \ac{spt} beams to get the expected signal profiles. The map noise assumed in constructing the filter includes the measured instrumental and atmospheric noise and sources of astrophysical noise, including the primary \ac{cmb}. Point sources are identified in a similar manner within each band independently, using only the instrument beams as the source profile \cite[][]{vieira10}.

Once \ac{sptsz} maps have been convolved with the multi-frequency matched filter, clusters are extracted with a simple peak-finding algorithm, with the primary observable $\xi$ defined as the maximum signal-to-noise of a given peak across a range of filter scales.  The \ac{sptsz} significance $\xi$ is a biased estimator that links to the underlying $\zeta$ as $\avg{\xi} = \sqrt{\zeta^2+3}$, because it is the maximum value identified through a search in sky position and filter angular scale \citep{vanderlinde10}. The observational scatter of $\xi$ around $\zeta$ is a unit-width Gaussian distribution corresponding to the underlying RMS noise of the \ac{sptsz} filtered maps.

In this work, we use the 90~GHz and 150 GHz  maps and employ the method described above to define an \ac{sptsz} significance for each X-ray selected cluster, but with two important differences: 1) We measure the \ac{sptsz} significance at the X-ray location, and 2) we use a cluster profile shape informed from the X-ray data.  We define this \ac{sptsz} significance as $\SN$, which is related to the unbiased \ac{sptsz} significance $\zeta$ as:
\begin{equation}\label{eq:xizeta}
  \zeta=\avg{\SN},
\end{equation}
where the angle brackets denote the average over many realizations of the experiment.  The observational scatter of $\SN$ around $\zeta$ is also a unit-width Gaussian distribution. Therefore $\SN$
is an unbiased estimator of $\zeta$, under the assumption that the true X-ray position and profile are identical to the true \ac{sze} position and profile -- a reasonable assumption -- given that both the X-ray and the \ac{sze} signatures are reflecting the intracluster medium properties of the clusters.  Note, however, that in the midst of a major merger the different density weighting of the X-ray and SZE signatures can lead to offsets \citep[][]{molnar12}.  

We model the relationship between $\zeta$ and the cluster mass through
\begin{equation}\label{eq:zetam}
  \zeta = \asz^\text{SPT}\Big(\frac{M_\text{500c}}{4.3\times10^{14} \msun} \Big)^{\bsz} \Big[ \frac{E(z)}{E(0.6)}\Big]^{\csz},
\end{equation}
where the intrinsic scatter on $\zeta$ is described by a log-normal distribution of width $\dsz$
\cite[B13;][]{reichardt13}.  We use $\asz^\text{SPT}$ to denote the amplitude of the original \ac{sptsz} scaling relation.  The differences in the depths of the \ac{sptsz} fields results in a re-scaling of the \ac{sptsz} cluster significance in spatially filtered maps.  For the field we study here, the relation requires a factor of 1.38 larger normalisation compared to the value in \cite{reichardt13}. 

For the massive \ac{sptsz} clusters (with $\xi>4.5$), the $\zeta$-mass relation is best parametrized as shown in \Fref{tab:xbcs-constraint} with $\csz=0.83\pm 0.30$ and $\dsz=0.21\pm0.09$ (B13).  In our analysis, we examine the characteristics of the lower mass clusters within the \ac{sptsz} survey.  To avoid a degeneracy between the scaling relation amplitude and slope, we shift the pivot mass to $1.5\times 10^{14}~\msun$, near the median mass of our sample and term the associated amplitude $\asz$.  At this pivot mass, with the normalisation factor mentioned previously, the equivalent amplitude parameter for the main \ac{sptsz} sample corresponds to $\asz=1.50$.  In \Fref{tab:xbcs-constraint} we also note the priors we adopt in our analysis of the low mass sample.  For our primary analysis we adopt flat priors on the amplitude and slope parameters and fix the redshift evolution and scatter at the values obtained by B13.

\subsection{Integrated  \texorpdfstring{$\ysz$}{Y500}}\label{sec:xbcs-yintegrated}

To facilitate the comparison of our sample with cluster physical properties reported in the literature, we also convert the $\SN$ to $\ysz$, which is the integration of the Compton-y parameter within a spherical volume with radius $\rfive$.  The central $y_{0}$ is
linearly linked to $\SN$ in the matched filter approach \citep{melin06}, with the corresponding Arnaud profile or $\beta$
profile as the cluster template.  The characteristic radii ($\rfive$ and $\rc$) are based on the X-ray measurements (S12), because the \ac{sze} observations are too noisy to constrain the profile accurately.

The projected circular $\beta$ profile for the filter is:
\begin{equation}
  \label{eq:betaprofile}
  y^\mr{(\beta)}_\text{cyl}(r) \propto (1-r^2/\rc^2)^{-(3\beta-1)/2},
\end{equation}
where $\beta$ is fixed to 1, consistent with higher signal to noise cluster studies \citep{plagge10}.  And the spherical $\ysz$ within the $\rfive$ is
\begin{equation}
  \label{eq:betay500}
  \ysz^\mr{(\beta)} = y_0 \times \pi \rc^{2} \ln(1+\rfive^{2}/\rc^{2}) \times f(\rfive/\rc),
\end{equation}
where $f(x)$ corrects the cylindrical result to the spherical value for the $\beta$ profile as:
\begin{equation}
  \label{eq:beta-deproject}
  f(x) = 2\frac{\ln(x+\sqrt{1+x^2})-x/\sqrt{1+x^2}}{\ln(1+x^2)}.
\end{equation}

The $\ysz^\text{(A)}$ for the Arnaud profile is calculated similarly except that the projected profile is calculated numerically within $5\rfive$ along the line-of-sight direction:
\begin{equation}\label{eq:arnaud-y-cyl}
  y^\text{(A)}_\text{cyl}(r) \propto
  \int_{-5\rfive}^{5\rfive} 
  P\Big(\frac{\sqrt{r^2+z^2}}{\rfive}\Big) \mathrm{d}z,
\end{equation}
where the pressure profile has the form
\begin{equation}
  P(x) \propto
  (c_\text{500}x)^{-\gamma_\text{A}} [1+(c_\text{500}x)^{\alpha_\text{A}}]^{(\gamma_\text{A}-\beta_\text{A})/\alpha_\text{A}},
\end{equation}
with $[c_\text{500},\gamma_\text{A},\alpha_\text{A},\beta_\text{A}]=[1.177,0.3081,1.0510,5.4905]$ \cite[]{arnaud10}.  The integration up to $5\rfive$ includes more than 99~per cent of the total pressure contribution.  The spherical $\ysz$ for the
Arnaud profile is:
\begin{equation}
  \label{eq:xbcs-yszproj-sph}
  \ysz^\text{(A)} = 2\pi y_{0} \int_0^{R_\text{500c}} y^\text{(A)}_\text{cyl}(r)
  r\mr{d}r /1.203,
\end{equation}
where the numerical factor 1.203 is the ratio between cylindrical integration and spherical integration for the adopted Arnaud profile parameters.

Measurements of $\ysz$ are sensitive to the assumed profile.  The Arnaud profile depends only on $\rfive$, while the $\beta$ profile depends on both $\rfive$ and $\rc$ and therefore $\ysz$ is sensitive to the ratio $\rc/\rfive$.  We find that with $\rc/\rfive=0.2$ the $\beta$ and Arnaud profiles provide $\ysz$ measurements in good agreement; this ratio is consistent with the previous \ac{sze} profile study using high mass clusters \citep{plagge10}.  Interestingly, the X-ray data indicate a  characteristic ratio of $0.11\pm0.03$ for our sample, and a shift in the $\rc/\rfive$ ratio from 0.2 to 0.1 leads to a $\sim$40~per cent decrease in $\ysz$.  Given that the \planck\ analysis to which we compare is carried out using the Arnaud profile, we adopt that profile for the analysis in Section~\ref{sec:xbcs-yszm} below.

The $\ysz$-mass scaling relation has been modelled using a representative local X-ray cluster sample \cite[]{arnaud10} and further studied in the \ac{sze} \cite[][P11]{andersson11} as
\begin{equation}\label{eq:xbcs-ym}
  \ysz = \ay \Big(\frac{M_{500}}{1.5\times10^{14} M_{\odot}}\Big)^{\by}E(z)^{2/3}\Big[\frac{D_\text{A}(z)}{500\text{Mpc}}\Big]^{-2},
\end{equation}
where $D_\text{A}(z)$ is the angular-diameter distance and the intrinsic scatter on $\ysz$ is described by a log-normal distribution of width $\sigma_\mr{\ln Y}=0.21$.  The observational scatter of $\ysz$ is propagated from the scatter of $\SN$.
In \Fref{sec:xbcs-result}, we fit this relation to the observations.

\section{Method}\label{sec:xbcs-method}
In this section, we describe the method we developed to fit the \ac{sze}-mass scaling relations of the low mass cluster population selected through the \ac{xbcs} and observed by the \ac{spt}.  In principle, we could use our cluster sample observed in X-ray and \ac{sze} to simultaneously constrain the cosmology and the scaling relations, in the so-called self-calibration approach \cite[]{majumdar04}.  However, self-calibration requires a large sample.  Without this, we take advantage of strong, existing cosmology constraints \citep[e.g.,][]{planck13-16,bocquet14} and knowledge of the $\lx$-mass scaling relation \citep[e.g.][]{pratt09}.  We focus only on the \ac{sze}-mass scaling relations, exploring the \ac{sze} characteristics of low mass galaxy clusters and groups.   In \Fref{sec:xbcs-lik} we present the method and in \Fref{sec:xbcs-submocks} we validate it using mock catalogs.

\begin{figure*}[htb]
  \centering
  \begin{subfigure}{0.3\textwidth}
    \hskip-0.75\textwidth \includegraphics[width=3.5in]{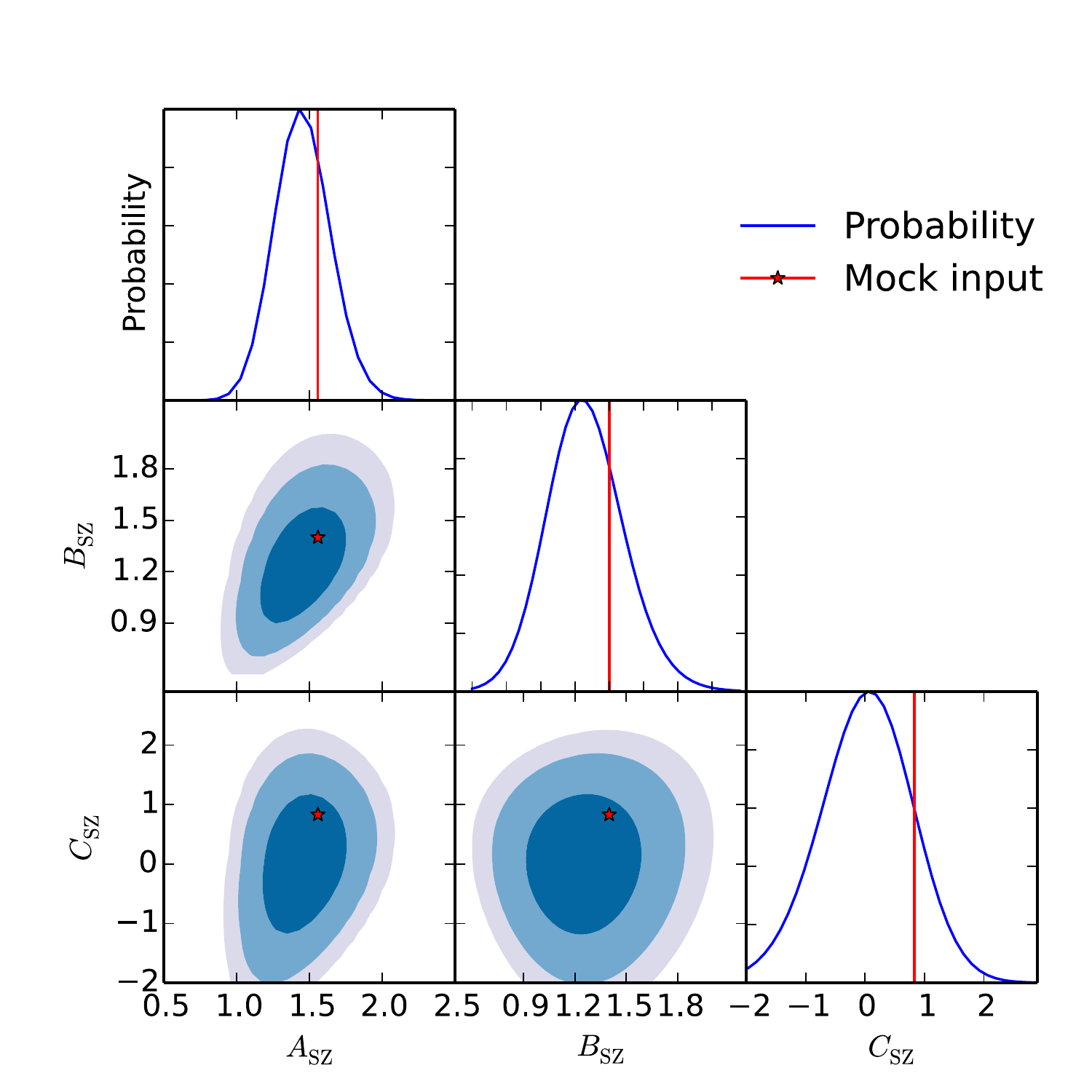}
  \end{subfigure}
  ~
  \begin{subfigure}{0.3\textwidth}
    \includegraphics[width=3.5in]{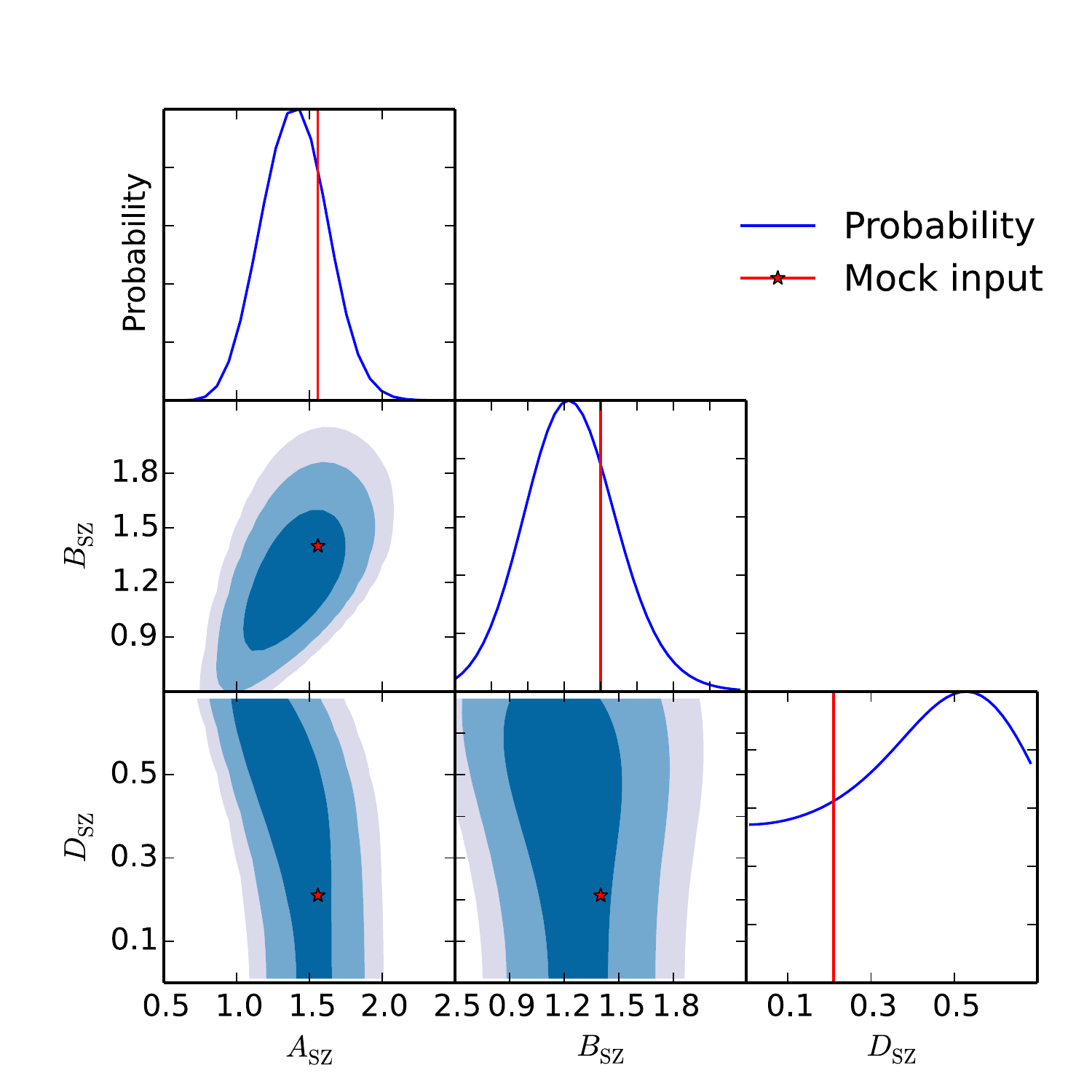}
  \end{subfigure}
  \caption[Constraints on the $\zeta$-mass relation from an analysis of the mock catalog]{Constraints on the $\zeta$-mass relation from an analysis of the mock catalog.  The left panel constrains $\asz$, $\bsz$, and $\csz$ with fixed $\dsz$.  And the right panel shows the result by fixing $\csz$ instead of $\dsz$.  The red lines and stars denote the input values of the scaling relation parameters of the mock catalog.  Histograms in each case show the recovered projected likelihood distribution for each parameter.  Joint constraints for different pairs of parameters are shown in blue with different shades indicating the 1, 2, and 3$\,\sigma$ levels.  }
  \label{fig:xbcs-mock-test}
\end{figure*}

\subsection{Description of the Method}\label{sec:xbcs-lik}

The selection biases on scaling relations include the Malmquist bias and the Eddington bias, 
which are manifestations of scatter and population variations associated with the selection observable.  Several methods have previously been developed (e.g., \citealt{vikhlinin09b, mantz10a, allen11}; B13; \citealt{bocquet14}) to account for the sampling biases when fitting scaling relation and cosmological parameters simultaneously.  In this analysis, we use a likelihood function that can be derived from the one presented in B13.  For a detailed discussion we refer the reader to Appendix~\ref{sec:likelihood};  here we present an overview of the key elements of this likelihood function.  

The likelihood function  $\mathcal{L}(\rsz)$ we use to constrain the \ac{sze}-mass relations is the product of the individual conditional probabilities to observe each cluster with \ac{sze} observable $Y_i$ (e.g., \ac{sptsz} significance $\SN$ or $\ysz$), given the cluster has been observed to have an X-ray observable $L_i$ and redshift $z_i$:
\begin{equation}\label{eq:xbcs-fullexp}
\mathcal{L}(\rsz)=\Pi_{i}\ P(Y_{i}|L_{i},z_{i},\bm{c}, \rx, \rsz, \tx),
\end{equation}
where $i$ runs over the cluster sample, $\rsz$ contains the parameters describing the \ac{sze} mass-observable scaling relation that we wish to study, $\bm c$ contains the cosmological parameters, $\rx$ contains the parameters describing the X-ray mass-observable scaling relation, and the survey selection in X-ray is encoded within $\tx$.  Note that the redshifts are assumed to be accurate such that the X-ray luminosity ($\lx$) is used instead of the true survey selection observable, which is the X-ray flux.  

As noted above, given the size of our dataset we adopt fixed cosmology $\bm c$ and X-ray scaling relation parameters $\rx$ to focus on the \ac{sze}-mass scaling relation.  In \Fref{sec:xbcs-result} we examine the sensitivity of our results to the current uncertainties in cosmology and the X-ray scaling relation and find them to be unimportant for our analysis.
Within this context, the conditional probability density function for cluster $i$ can be written as the ratio of the expected number of clusters $\mr{d}N$ with observables $Y_i$, $L_i$ and $z_i$ within infinitesimal volumes $\mr{d}Y$, $\mr{d}L$ and $\mr{d}z$:
\begin{equation}\label{eq:xbcs-single-like}
P(Y_{i}|L_{i},z_{i},\rsz,\tx) =\frac{\mr{d}N(Y_{i}, L_{i}, z_{i}|\rsz,\tx)}{\mr{d}N(L_{i}, z_{i}|\tx)},
\end{equation}
where we have dropped the cosmology $\bm{c}$ and X-ray scaling relation parameters $\rx$ because they are held constant.  Typically, the survey selection $\tx$ is a complex function of the redshift and X-ray flux, but in the above expression it is simply the probability that a cluster with X-ray luminosity $L_i$ and redshift $z_i$ is observed 
(i.e. $\mr{d}N\left(Y_{i},L_{i},z_{i}|\rsz,\tx\right)=\tx(L_{i},z_{i})\mr{d}N(Y_{i},L_{i},z_{i}|\rsz)$); 
in \Fref{eq:xbcs-single-like} this same factor appears in both the numerator and denominator, and therefore it cancels out.   Thus, studying the \ac{sze} properties of an X-ray selected sample does not require detailed modelling of the selection.  If the selection were based on both $L$ and $Y$, then there would be no cancellation, because the selection probability in the numerator would be just $\Theta(L_{i},Y_{i},z_{i})$ while in the denominator it would have to be marginalised over the unobserved $Y$ as $\int \Theta(Y,L_{i},z_{i})\mr{d}Y$  (see \Fref{eq:xbcs-selection-yl}).

With knowledge of the cosmologically dependent mass function $n(M,z)\equiv\mr{d}N(M,z|\bm{c})/\mr{d}M\mr{d}z$ \cite[]{tinker08}, the ratio of the expected number of clusters can be written as: 
\begin{equation}\label{eq:xbcs-simplify-likelihood}
P(Y_{i}|L_{i},z_{i},\rsz) =\frac{\int \mr{d}M P(Y_{i},L_{i}|M,z_{i},\rsz)\,n(M,z_{i})}{\int \mr{d}M P(L_{i}|M,z_{i})\,n(M,z_{i})}.
\end{equation}

We emphasise that there is a residual dependence on the X-ray selection in our analysis in the sense that we can only study the \ac{sze} properties of the clusters that have sufficient X-ray luminosity to have made it into the sample.  This effectively limits the mass range over which we can use the X-ray selected sample to study the \ac{sze} properties of the clusters.

To constrain the scaling relation in the presence of both observational uncertainties and intrinsic scatter, we further expand the conditional probability density functions in Equation~(\ref{eq:xbcs-simplify-likelihood}):
\begin{align}
P(Y_{i},L_{i}|M,z_{i},\rsz) = \iint&\mr{d}Y_\mr{t} \mr{d}L_\mr{t}\ P(Y_{i}, L_{i}|Y_\mr{t},L_\mr{t}) \nonumber\\
&\times P(Y_\mr{t},L_\mr{t} |M,z_{i},\rsz) , \label{eq:xbcs-ymlcore} \\
P(L_{i}|M,z_{i}) = \int&\mr{d}L_\mr{t}\ P(L|L_\mr{t}) P(L_\mr{t}|M,z_{i}),\label{eq:xbcs-mlcore}
\end{align}
where, as above, $Y_{i}$ and $L_{i}$ are the observed values, and $Y_\mr{t}$ and $L_\mr{t}$ are the true underlying observables related to mass through scaling relations that have intrinsic scatter.  The first factor in each integral represents the measurement error, and the second factor describes the relationship between the pristine observables and the halo mass. Improved data quality affects the first factor, but cluster physics dictates the form of the second.  These second factors are fully described by the power law mass-observable relations in Equations (\ref{eq:xbcs-lm}), (\ref{eq:zetam}), and (\ref{eq:xbcs-ym}) together with the adopted log-normal scatter.

We use this likelihood function under the assumption that there is no correlated scatter in the observables; in \Fref{sec:xbcs-submocks} we use mock samples that include correlated scatter to examine the impact on our results. 

\subsection{Validation with Mock Cluster Catalogs}\label{sec:xbcs-submocks}

We use mock samples of clusters to validate our likelihood and fitting approach and to explore our ability to constrain different parameters. Specifically, we generate ten larger mock surveys of 60~$\sdeg$, with a similar flux limit of $1\times 10^{-14} \text{erg s}^{-1}\text{cm}^{2}$ and $z>0.2$.  Each mock catalog contains $\sim400$ clusters, or approximately eight times as many as in the observed sample. The $\SN$ of the sample spans $-2.2\le\SN\le7.8$ with a median value of $1.4$.  We include both the intrinsic scatter and observational uncertainties for both the $\lx$ and the $\SN$ in the mock catalog.  The intrinsic scatter is lognormal distributed with values given as $\sigma_{\ln\lx}$ ($\dsz$).  The observational uncertainties in $\lx$ and $\SN$ are modelled as normal distributions. The standard deviation used for $\lx$ is proportional to $\sqrt{\lx}$ to mimic the Poisson distribution of photon counts, while the standard deviation for $\SN$ is 1.

Here we focus on recovering the four \ac{sptsz} $\zeta$-mass relation parameters from the mock catalog; the fiducial values for these parameters are the B13 best-fitting values.  We scan through the parameter space using a fixed grid. The following results contain 41 bins in each parameter direction. Given the limited constraining power, we validate the parameters using two different sets of priors. In the first set we adopt flat priors on $\asz$, $\bsz$, and $\csz$ with fixed $\dsz$. In the second set we adopt flat positive priors on $\asz$, $\bsz$, and $\dsz$ with fixed $\csz$. All other relevant parameters are fixed, including the $\lx$-mass scaling and the cosmological model.  

Our tests show good performance of the method.  Using ten mock samples that are each ten times larger than our observed sample, and fitting for 3 parameters in each mock, we recover the parameters to within the marginalised $1\,\sigma$ statistical uncertainty 70~per cent of the time and to within $2\,\sigma$ for the rest.  \Fref{fig:xbcs-mock-test} illustrates our $\zeta$-mass parameter constraints from one mock sample.  Note that the constraints on $\csz$ and $\dsz$ are both weak and exhibit no significant degeneracy with the other two \ac{sptsz} scaling parameters. We take this as motivation to fix $\csz$ and $\dsz$ and focus on the amplitude $\asz$ and slope $\bsz$ in the analysis of the observed sample.  We have repeated this testing in the case of the $\ysz$-mass relation, and we see no difference in behavior.  

We also investigate the sensitivity of our method when a correlation between intrinsic scatter in the X-ray and $\SN$ is included.  Cluster observables can be correlated through an analysis approach.  For example, if one uses the $\lx$ as a virial mass estimate, then when $\lx$ scatters up by 40~per cent, it leads to a 5~per cent increase in radius, and 8~per cent increase in $\ysz$ if the underlying SZE brightness distribution is described by the \cite{arnaud10} profile.  In comparison, the intrinsic scatter of $\ysz$ about mass is about 20~per cent, which in this example would still dominate over the correlated component of the scatter.  Correlated scatter in different observable-mass relations can also reflect underlying physical properties of the cluster that impact the two observables in a similar manner.

We find that even with a correlation coefficient $\rho=0.5$ between the intrinsic scatter of the two observables, the change in constraints extracted using a no correlation assumption is small.  Thus, our approximation does not lead to significant bias in the analysis of this sample.   This result is also consistent with the fact that by extending Equations~(\ref{eq:xbcs-ymlcore}) and (\ref{eq:xbcs-mlcore}) to include multi-dimensional log-normal scatter distributions, we find the constraint on correlated scatter in the mock catalog to be very weak.  We therefore do not include the possibility of correlated scatter when studying the real sample.

\section{Results}\label{sec:xbcs-result}

In this section, we present the observed relationship between the \ac{sze} significance $\SN$ at the position of the X-ray selected cluster and the predicted value given the measured X-ray luminosity of the system.  Thereafter, we test -- and rule out -- the null hypothesis that the \ac{sze} signal at the locations of the X-ray selected clusters is consistent with noise. We then present constraints on the \ac{sptsz} $\zeta$-mass and $\ysz$-mass relations.  We end with a discussion of possible systematics and a presentation of the point source population for this X-ray selected group and cluster sample.

\subsection{SPT Significance Extraction}

We extract the $\SN$ from the \ac{sptsz} multi-frequency-filtered map at the location of each \ac{xbcs} selected cluster as described in \Fref{sec:xbcs-data-description}. In the primary analysis, we adopt
three matched-filtered maps from the \ac{sptsz} data, one each for $\beta$-model profiles with $R_\text{c} =$ 0.25, 0.5, and 0.75 arcmin, and we extract the value of $\SN$ for each cluster from the map that most closely matches the X-ray-derived $R_\text{c}$ value for that cluster. The $\SN$ is extracted at the X-ray-derived cluster position.  The measured $\SN$ values are presented in \Fref{tab:xbcs-cat}.  We have also tried extracting \ac{sptsz} significance by making a matched-filtered map for every cluster, using a filter with the exact X-ray-derived value of $R_\text{c}$, and the change in the results is negligible. 

We have also investigated the dependence of $\SN$ on the assumed cluster profile. We repeated the analysis described above using the Arnaud profile and a $\beta$ profile with $\beta=2/3$.  The resulting changes in the extracted values of $\SN$ are less than 3~per cent of the measurement uncertainty on the individual $\SN$ values.  A similar lack of sensitivity to the assumed cluster profile is seen in the $\xi > 5$ \ac{sptsz} derived cluster samples.

\begin{figure}
  \hskip-0.0in\includegraphics[width=3.3in]{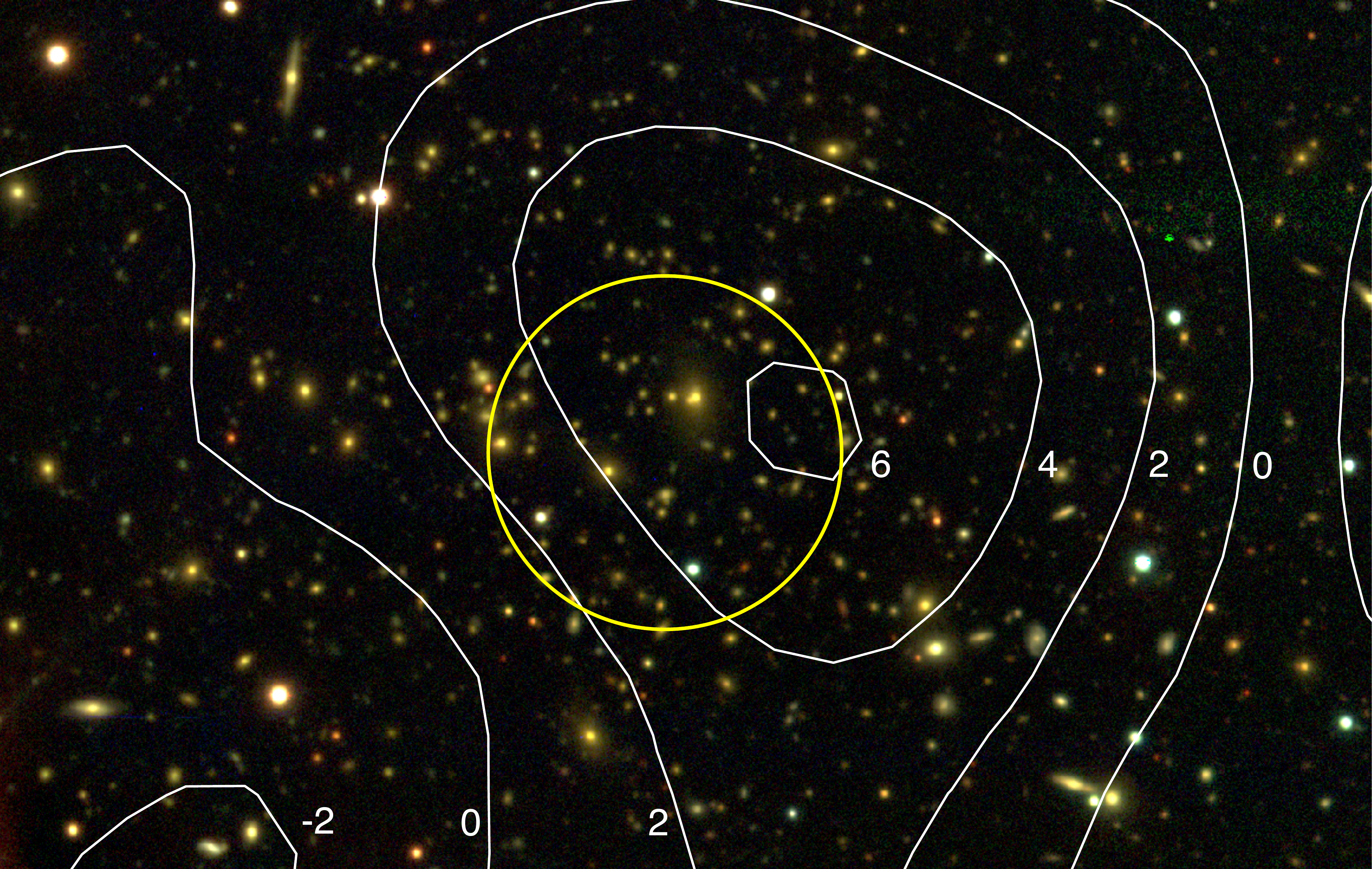}
  \caption[BCS optical image of cluster 044.]
  {\ac{bcs} optical pseudo-colour image of cluster 044 in {\it gri} bands. The yellow circle (1.5~arcmin diameter) centred at the X-ray peak indicates the rough size of the \ac{spt} beam (1.2~arcmin FWHM in 150~GHz and 1.6~arcmin in 95~GHz).  The \ac{sptsz} filtered map is overlaid with white contours, which are marked with the significance levels.  The offset between the X-ray centre and the \ac{sze} peak is $0.75$\,arcmin, and the BCG for this system lies near those two centres.}
  \label{fig:xbcs-clu044}
\end{figure}

The cluster with the strongest detection in the \ac{sptsz} maps is illustrated in \Fref{fig:xbcs-clu044}, which contains a pseudo-colour optical image with %
\ac{sptsz} signal-to-noise contours in white.  The \ac{sptsz} significance, $\xi$, of this cluster is 6.23 corresponding to maximum signal-to-noise in the filtered map (SPT-CLJ2316-5453, Bleem et al. in prep.), whereas the $\SN$ is 4.58 at the X-ray position with $\rc$ of 0.367~arcmin.  This reduction in signal to noise is expected because there is noise in the \ac{sze} map, and the \ac{sptsz} cluster is selected to lie at the peak $\xi$.

\begin{table*}
  \begin{center} 
    \caption[SPT significance of the XMM-BCS sample.]{\ac{sptsz} $\SN$ of \ac{xbcs} sample.}\label{tab:xbcs-cat}
    \begin{tabular}{|c|r|r|r|r|r|r|c|c|} 
      \hline 
ID & \begin{tabular}{@{}c@{}} $L_{\mathrm{X,500,bol}}$ \\
  $[10^{42}~\text{erg/s}]$\end{tabular} & 
\begin{tabular}{@{}c@{}} $\Delta L_{\mathrm{X,500,bol}}$ \\
  $[10^{42}~\text{erg/s}]$ \end{tabular} &
Redshift &\begin{tabular}{@{}c@{}} Redshift \\
  uncertainty  \end{tabular} & \begin{tabular}{@{}c@{}} $R_\text{c}$ \\
  $[\text{arcmin}]$ \end{tabular} & 
$\SN$  &\begin{tabular}{@{}c@{}} SPT point source \\ separation [arcmin] and SN\end{tabular} &
\begin{tabular}{@{}c@{}} SUMSS point source \\ separation
[arcmin] \end{tabular} \\  \hline
011 & $   345.2$&$    51.6$ & $   0.97$&$   0.10$ &   0.185 &    0.99& - & - \\
018 & $    66.3$&$     6.5$ & $   0.39$&$   0.04$ &   0.239 &    1.90& - &   0.92  \\
032 & $   684.0$&$    56.8$ & $   0.83$&$   0.07$ &   0.272 &    3.04& - &   1.70, 2.30, 3.97  \\
033 & $   209.0$&$    17.6$ & $   0.79$&$   0.05$ &   0.189 &    2.34& - & - \\
034 & $    16.0$&$     2.5$ & $   0.28$&$   0.02$ &   0.197 &   -0.38& - & - \\
035 & $    91.0$&$    14.3$ & $   0.67$&$   0.05$ &   0.164 &    2.78& - &   0.10, 1.56  \\
038 & $    16.3$&$     2.5$ & $   0.39$&$   0.05$ &   0.147 &   -0.20& - &   1.85  \\
039 & $    19.4$&$     1.2$ & $   0.18$&$   0.04$ &   0.315 &   -0.34& - &   2.91  \\
044 & $   310.5$&$    20.5$ & $   0.44$&$   0.02$ &   0.367 &    4.58& 3.87\quad 4.84  &   0.22  \\
069 & $   124.9$&$    21.5$ & $   0.75$&$   0.07$ &   0.165 &    1.38& 3.40\quad  6.34&   3.42  \\
070 & $   137.9$&$     2.8$ & $   0.152$&$   0.001$ &   0.726 &    1.80& - & - \\
081 & $    93.1$&$    15.4$ & $   0.85$&$   0.12$ &   0.133 &   -1.56& - & - \\
082 & $    53.6$&$     9.2$ & $   0.63$&$   0.05$ &   0.144 &    0.55& - & - \\
088 & $   122.1$&$    16.7$ & $   0.43$&$   0.04$ &   0.271 &   -0.10& - &   2.96  \\
090 & $    25.4$&$     5.8$ & $   0.58$&$   0.02$ &   0.120 &    0.30& - & - \\
094 & $    26.3$&$     2.9$ & $   0.269$&$   0.001$ &   0.243 &    2.20& - &   1.48  \\
109 & $   196.9$&$    28.8$ & $   1.02$&$   0.09$ &   0.145 &    1.09& - &   0.19  \\
110 & $    68.8$&$     9.3$ & $   0.47$&$   0.06$ &   0.205 &   -1.07& - &   0.10  \\
126 & $    82.0$&$     6.1$ & $   0.42$&$   0.02$ &   0.240 &    0.03& - &   1.22  \\
127 & $     8.4$&$     1.0$ & $   0.207$&$   0.001$ &   0.207 &    1.28& - & - \\
132 & $   319.3$&$    35.7$ & $   0.96$&$   0.17$ &   0.182 &    1.74& - & - \\
136 & $    86.8$&$     7.3$ & $   0.36$&$   0.02$ &   0.282 &   -3.58&  1.11\quad 5.84  &   1.00  \\
139 & $     8.7$&$     1.2$ & $   0.169$&$   0.001$ &   0.252 &   -0.17& - &   0.44  \\
150 & $    37.7$&$     1.8$ & $   0.176$&$   0.001$ &   0.403 &   -3.34& 0.13\quad 4.23 &   0.05, 2.29  \\
152 & $     3.4$&$     0.6$ & $   0.139$&$   0.001$ &   0.219 &   -0.45& - & - \\
156 & $   166.0$&$    11.7$ & $   0.67$&$   0.06$ &   0.202 &    3.01& - & - \\
158 & $   104.2$&$    15.6$ & $   0.55$&$   0.03$ &   0.205 &    1.94& - & - \\
210 & $    45.0$&$     9.0$ & $   0.83$&$   0.09$ &   0.105 &    0.18& - & - \\
227 & $    14.5$&$     1.8$ & $   0.346$&$   0.001$ &   0.157 &   -1.03& - & 0.06\\
245 & $    38.1$&$     7.1$ & $   0.62$&$   0.03$ &   0.130 &    0.24& - &   1.38  \\
275 & $    17.8$&$     2.7$ & $   0.29$&$   0.03$ &   0.198 &   -0.46& - &   2.12  \\
287 & $    31.1$&$    11.0$ & $   0.57$&$   0.04$ &   0.131 &   -0.02& - & - \\
288 & $    89.0$&$    17.4$ & $   0.60$&$   0.04$ &   0.180 &   -0.25& - &   0.62  \\
357 & $    66.3$&$     8.3$ & $   0.48$&$   0.06$ &   0.198 &   -0.97& - & - \\
386 & $    17.7$&$     4.8$ & $   0.53$&$   0.05$ &   0.115 &    0.83& 0.417\quad 4.53$^*$ & - \\
430 & $     4.5$&$     0.9$ & $   0.206$&$   0.001$ &   0.167 &   -0.67& - & - \\
444 & $    69.1$&$    13.8$ & $   0.71$&$   0.05$ &   0.141 &   -0.13& - & - \\
457 & $     1.1$&$     0.3$ & $   0.100$&$   0.001$ &   0.201 &   -1.24& - & - \\
476 & $     6.2$&$     0.7$ & $   0.101$&$   0.001$ &   0.365 &   -0.12& - &   1.03  \\
502 & $    47.2$&$     4.2$ & $   0.55$&$   0.05$ &   0.156 &   -0.30& - & - \\
511 & $    23.4$&$     3.7$ & $   0.269$&$   0.001$ &   0.233 &    0.11& - &   0.15, 2.37 \\
527 & $   160.8$&$    26.2$ & $   0.79$&$   0.06$ &   0.172 &    0.83& - &   3.96  \\
528 & $     6.4$&$     2.1$ & $   0.35$&$   0.02$ &   0.117 &    0.57& - & - \\
538 & $     5.1$&$     2.1$ & $   0.20$&$   0.02$ &   0.179 &    0.30& - & - \\
543 & $   134.5$&$    29.6$ & $   0.57$&$   0.03$ &   0.217 &    1.10& - & - \\
547 & $     4.1$&$     1.3$ & $   0.241$&$   0.001$ &   0.140 &   -6.45& 0.20\quad 6.75 &   0.12, 2.89  \\
      \hline 
    \end{tabular}
  \end{center}
 \raggedright
$^*$Detected in 220~GHz.
\end{table*}%

\subsection{Testing the Null Hypothesis}

To gain a sense of the strength of the \ac{sze} detection of the ensemble of \ac{xbcs} clusters, we test the measured significance around \ac{sze} null positions.  A single null catalog consists of the same number of clusters as the \ac{xbcs} sample where the X-ray luminosities and redshifts are maintained, but the \ac{sptsz} significances $\SN$ are measured at random positions.  We then carry out a likelihood analysis of three null catalogs.  When fixing the slope $\bsz$ of the scaling relation, we find that the normalisation factor $\asz$ is constrained to be $<0.56$ at 99~per cent confidence level for all three null samples we tested.  Because this constraint on the amplitude is small compared to the expected normalisation for the \ac{xbcs} sample, we have essentially shown that there should be sufficient signal to noise to detect the \ac{sze} signature of the cluster ensemble.

\subsection{SPT \texorpdfstring{$\zeta$}{zeta}-mass Relation}
\label{sec:xbcs-xim}

We explore the \ac{sze} signature of low mass clusters by constraining the $\asz$ and $\bsz$ parameters with the approach described and tested above.  The X-ray luminosity-mass scaling relation, \Fref{eq:xbcs-lm}, is directly adopted with the additional observational uncertainties of each cluster that are listed in \Fref{tab:xbcs-cat} (bolometric luminosities presented in S12).

We present results for four different subsets of our sample: 1) the full sample without removal of any cluster; 2) the sample excluding any cluster with a point source detected  at $>$4\,$\sigma$ in any \ac{spt} observing band within a 4~arcmin radius of the X-ray cluster (see \Fref{tab:xbcs-cat}), hereafter SPT-NPS sample; 3) the SPT-NPS clusters with redshift larger than 0.3, hereafter SPT-NPS($z>0.3$), which is the best match to the selection of the \ac{sptsz} high mass sample in B13 and 4) the sample without any \acl{sumss} \cite[\acsu{sumss},][]{bock99,mauch03} point sources  in 4~arcmin radius.  We discuss further the astrophysical nature and impact of point sources in \Fref{sec:xbcs-PScontamination}.

\begin{figure}
  \hskip-0.2in\includegraphics[width=3.5in]{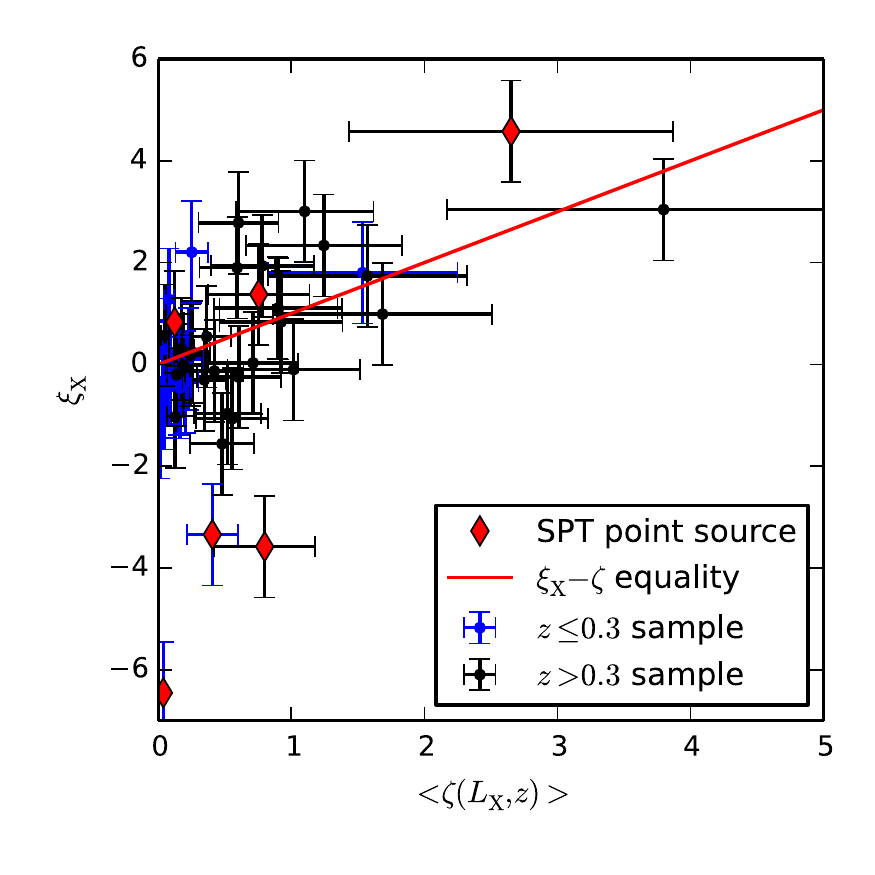}
  \caption[The measured significance $\SN$ versus the expected \ac{sptsz} $\avg{\zeta(\lx,z)}$ from XMM-BCS sample]{The measured significance $\SN$ versus the expected \ac{sptsz} $\avg{\zeta(\lx,z)}$, where the best-fitting relation from the SPT-NPS sample and sampling bias corrections are applied.  Overplotted is the line of equality.  Clusters close to \ac{spt} point sources are marked with red diamonds.}
  \label{fig:xbcs-corrbias}
\end{figure}

In \Fref{fig:xbcs-corrbias}, we illustrate the $\zeta$-mass relation obtained by plotting the observed $\SN$ versus the expected $\avg{\zeta(\lx,z)}$, estimated using \Fref{eq:xbcs-fullexp}.  Here we use the best fit scaling relation from the SPT-NPS (black points only).  Note that the typical bias correction on the mass is about 10 percent at the high mass end.  

\begin{figure}
  \hskip-0.3in\includegraphics[width=3.5in]{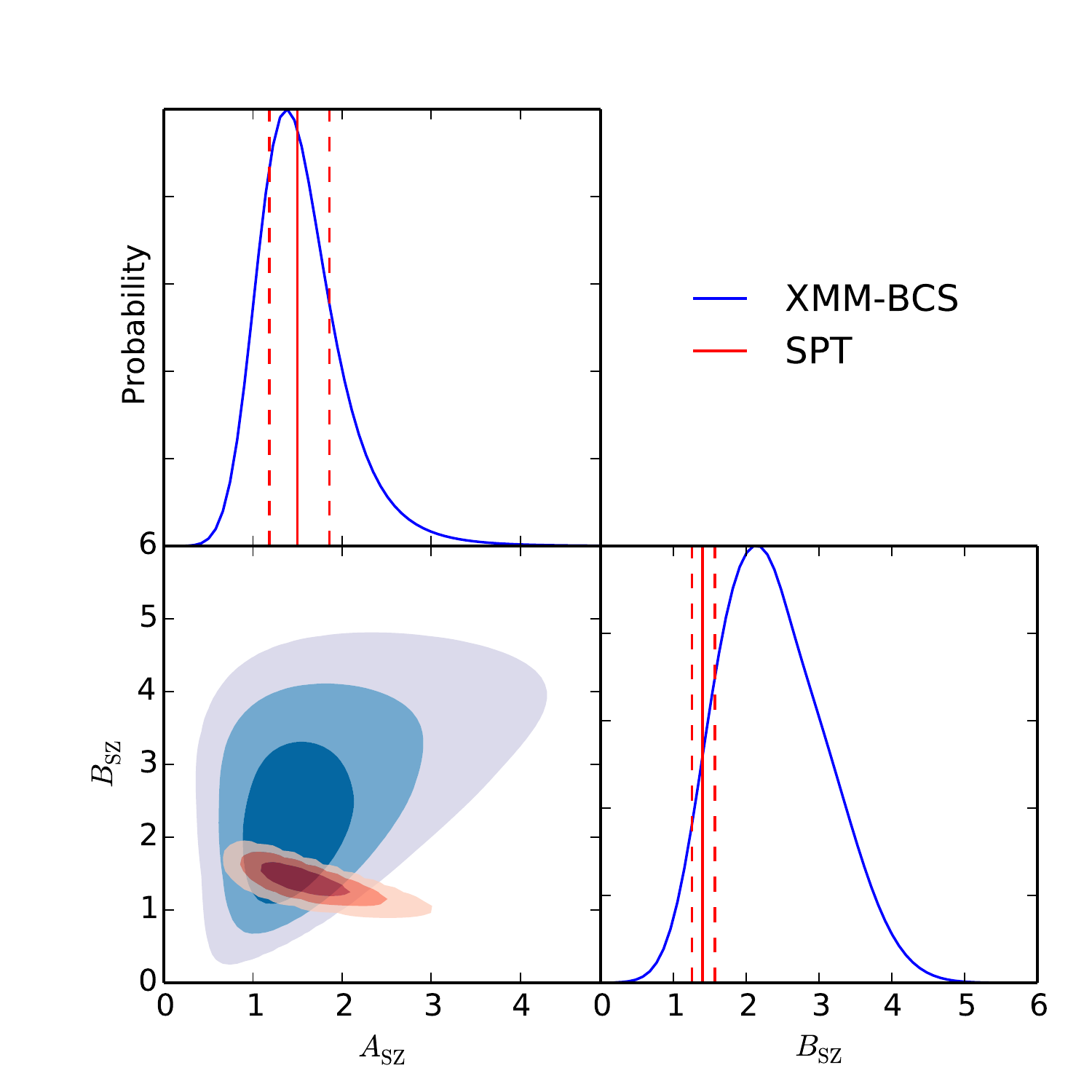}
  \caption[Constraints on the SPT $\zeta$-mass relation parameters $\asz$ and $\bsz$ for the non-point source sample (SPT-NPS)]{Constraints on the \ac{sptsz} $\zeta$-mass relation parameters $\asz$ and $\bsz$ for the non-point source sample (SPT-NPS).  The different shading indicates 1, 2, and 3$\sigma$ confidence regions.  The constraints from the \ac{sptsz} high mass clusters (B13) are shown in red with 68~per cent confidence regions marked with dashed lines.  The amplitudes for low and high mass clusters are compatible, but the slope is higher for low mass systems by about 1.4$\,\sigma$.}
  \label{fig:xbcs-AszBsz}
\end{figure}

We explore the likelihood as a function of $\asz$ and $\bsz$ and show the parameter constraints for the three samples in \Fref{tab:xbcs-params}, and we show the likelihood distribution of the SPT-NPS sample in \Fref{fig:xbcs-AszBsz}. We also show marginalised single parameter probability distributions, which we use to calculate the 68~per cent confidence region for each parameter.  This confidence region along with the modal value is reported in \Fref{tab:xbcs-params}. For comparison, the constraints from the B13 analysis are shown in red.

\begin{table}
  \begin{center}
    \caption[Constraints on the SPT zeta-mass relation
    parameters.]{Constraints on the SZE $\zeta$-mass relation
      parameters.\label{tab:xbcs-params}}
    \begin{tabular}{|c|c|c|}
      \hline
      & $\asz$ & $\bsz$ \\ \hline
      SPT High Mass (B13) & $1.50\pm0.34$& $1.40\pm0.16$\\
      Prior & $[0.1-5]$ & $[0.1-6]$ \\
            Full sample & $1.38^{+0.46}_{-0.36}$ & $2.80^{+0.66}_{-0.63}$\\
      SPT-NPS & $1.37^{+0.48}_{-0.38}$&$2.14^{+0.86}_{-0.66}$ \\
      SPT-NPS ($z>0.3$) &$1.37^{+0.60}_{-0.46}$ & $2.31^{+1.31}_{-0.86}$\\
      SPT-No-SUMSS & $1.42^{+0.58}_{-0.43}$ & $2.14^{+0.91}_{-0.71}$ \\
      \hline
    \end{tabular}
    \label{tab:xbcs-constraint}
  \end{center}
\end{table}

All three low mass subsamples show similar normalisation to the extrapolated high mass \ac{sptsz} sample, but there is a preference for larger slopes.  The SPT-NPS sample is the best for comparison to the \ac{sptsz} high mass sample used in B13; this is because the \ac{spt} point sources have been removed to mimic the \ac{spt} cluster catalog selection and because there is no measurable difference between the SPT-NPS samples with or without the redshift cut.  

The fact that we find consistent results with or without a low-redshift cut may at first be surprising, given that analyses of the high-mass \ac{sptsz} cut all clusters below z=0.3.  In the \ac{sptsz} high mass sample, the low redshift clusters are cut because the angular scales of these clusters begin to overlap the scales where there is significant \ac{cmb} primary anisotropy, making extraction with the matched filter approach using two frequencies difficult.  However the \ac{xbcs} clusters are low mass systems with corresponding $\rc$ less than 1~arcmin even at low redshift.  So we are able to recover the same scaling relation with or without the low redshift clusters.

The fully marginalised posterior probability distributions for $\bsz$ can be used to quantify consistency between the two datasets.  We do this for any pair of the distributions $P_i\left(\theta\right)$ by first calculating the probability density distribution of the difference $\Delta \theta$: 
\begin{equation}
P(\Delta \theta) = \int \mr{d}\theta P_{1}(\theta)P_{2}(\theta-\Delta\theta).
\label{eq:xbcs-1dp2exceed}
\end{equation} 
We then calculate the likelihood $p$ that the origin ($\Delta\theta=0$) lies within this distribution as
\begin{equation}
p = \int_S \mr{d}\Delta\theta\  P(\Delta\theta)
\end{equation}
where $S$ is the space where $P(\Delta\theta)<P(\Delta\theta=0)$.  We then convert this $p$ value to an equivalent $N$-$\sigma$ significance within a normal distribution. 

Overall, there is no strong statistical evidence that the low mass clusters behave differently than expected by simply extrapolating the high mass scaling relation to low mass; the slope parameter $\bsz$ of the \ac{sptsz} high mass and SPT-NPS samples differs by only $1.4\,\sigma$ (\Fref{tab:xbcs-constraint}).  The full sample has a $2.6\,\sigma$ higher $\bsz$ than the \ac{sptsz} high mass sample \citep{benson13}.  This steeper slope is presumably due to the contaminating effects of the \ac{spt} point sources. We find three outliers below the $\lx\text{-}\SN$ distribution (\Fref{fig:xbcs-corrbias}) that are all contaminated by \ac{spt} point sources. We list the separation between the cluster centres and the nearest \ac{spt} point source in \Fref{tab:xbcs-cat}.

It is clear from \Fref{fig:xbcs-corrbias}  and from the results for the full sample
that including X-ray-selected clusters that are associated with point
sources that are independently detected in \ac{sptsz} data can bias the
derived \ac{sze}-mass relation. In these cases, the affected clusters can
be removed from the sample, and this particular bias can be easily avoided.
Point sources that are not detected in the \ac{sptsz} data but which could be
significantly affecting the measured \ac{sze} signal -- particularly in low-mass
clusters and groups -- do remain a potential issue. We discuss this and the effect 
of point sources on our results more generally in \Fref{sec:xbcs-PScontamination}.

In addition to the X-ray bolometric luminosities, we test the luminosities based on two other bands (0.5--2.0~keV and 0.1--2.4~keV) as predictors of the cluster mass.  After applying the appropriate $\lx$-mass relations listed in \Fref{tab:xbcs-lxm} we find that the changes to the parameter estimates are small.  The largest change is on the slope of the \ac{sptsz} $\zeta$-mass relation, but the difference is less than $0.2\,\sigma$.  Thus, the choice of X-ray luminosity band is not important to our analysis.

Our results show some dependence on the assumed $\lx$-mass scaling relation.  Adopting the \citet{vikhlinin09b} scaling relation has no significant impact on our results.  However, with the \citet{mantz10b} $\lx$-mass relation, the slope decreases to $\bsz\sim1.57$ from 2.14, which makes the SPT-NPS sample almost a perfect match to the high mass \ac{sptsz} scaling relation.  This shift is not surprising, because the \citet{mantz10b} $\lx$-mass relation has a very different slope from \citet{pratt09} (1.63 vs. 2.08, respectively).  This causes clusters with a $\lx<1\times10^{44}~\mathrm{erg\ s}^{-1}$to have significantly lower estimated masses when assuming the \citet{mantz10b} relation (20~per cent on average and $\sim40$~per cent at the low mass end).  We expect the \citet{pratt09} relation to be more appropriate for our analysis, because the \citet{mantz10b} relation was calibrated from higher mass clusters, using only clusters with $\lx >2.5\times10^{44}~\mathrm{erg\ s}^{-1}$, above the majority of \ac{xbcs} clusters.  Also we note the change of $\SN$ caused by the updated $\rfive(\lx)$ is negligible, which has been shown also in \citet{saliwanchik13}.
  
\begin{figure}
  \hskip-0.3in\includegraphics[width=3.5in]{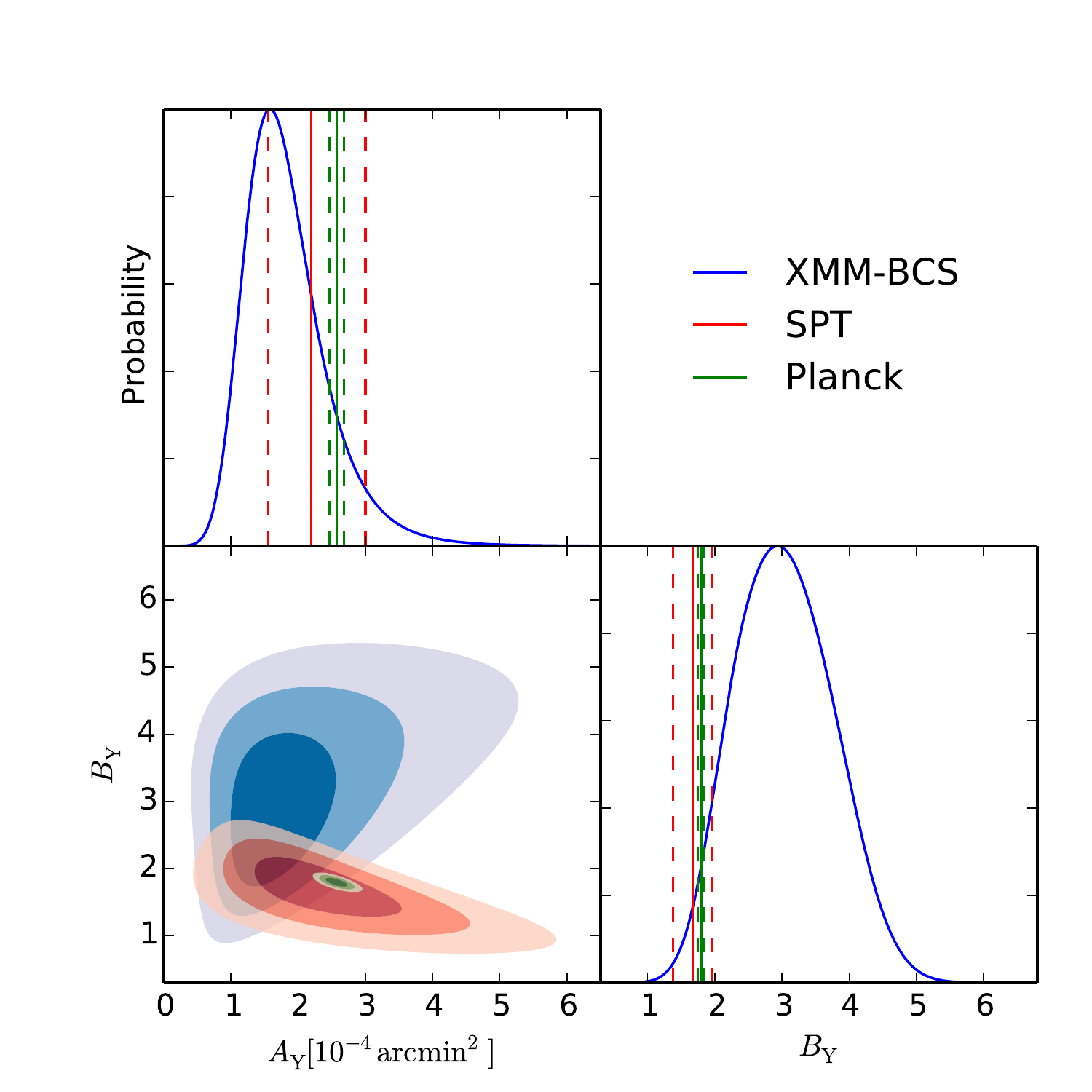}
  \caption[Constraints on the $\ysz$-mass relation parameters $\ay$ and $\by$ for the non-point source sample (SPT-NPS)]{Constraints on the $\ysz$-mass relation parameters $\ay$ and $\by$ for the non-point source sample (SPT-NPS).  The SPT-NPS constraints are shown in blue and different shades show the 1, 2, and 3\,$\sigma$ levels.  The red is for the \ac{sptsz} result \protect\citep{andersson11}, and the green is the best fit from the \planck\ analysis (\plx).  Marginalised constraints for each parameter are shown in blue with best fit and $1\,\sigma$ confidence regions marked by solid and dashed lines, respectively.}
\label{fig:xbcs-ym}
\end{figure}

\subsection{SZE \texorpdfstring{$\ysz$}{Y500}-mass
  Relation}\label{sec:xbcs-yszm}

We measure the $\ysz$-mass relation, using the SPT-NPS sample.  A similar fitting approach is used to account for the selection bias and with the same shifted pivot mass in \Fref{eq:xbcs-ym} of $1.5\times10^{14}~\msun$.  The best fit parameters and uncertainties are presented in \Fref{tab:xbcs-yszm} along with the results from \cite{andersson11} and \plx, which are adjusted to use our lower pivot mass. The $\ysz$ is based on the Arnaud profile and the $\lx$ is based on the X-ray luminosity measured within the 0.1--2.4~keV band, which facilitates the comparison with the \plx\ result.  The impact from different profiles is discussed later in this section.

\Fref{fig:xbcs-ym} shows the joint parameter and fully marginalised constraints for $\ay$ and $\by$.  The shaded regions denote the 1, 2, and 3\,$\sigma$ confidence regions as in \Fref{fig:xbcs-AszBsz} with blue for the SPT-NPS, red for the \ac{sptsz} sample \cite[]{andersson11}, and green for the \planck\ sample (\plx).  This figure shows that the low mass SPT-NPS sample has rather weak constraints that are shifted with respect to the high mass \ac{sptsz} sample and the \planck\ sample.  

We estimate the significance of the difference using the method described in \Fref{sec:xbcs-xim}.  We quantify the consistency between any pair of the two-parameter distributions $P_i\left(\bm\theta\right)$ by calculating a $p$ value in a manner similar to that in \Fref{eq:xbcs-1dp2exceed} with the null hypothesis $\Delta \bm \theta =0$.  Using this approach, we calculate that the SPT-NPS sample is roughly consistent with the high mass \ac{sptsz} sample (a $1.4\,\sigma$ difference) but is in tension with the \planck\ result (a $2.8\,\sigma$ difference).

Also shown in \Fref{fig:xbcs-ym} are the fully marginalised single parameter constraints.  These distributions indicate that the normalisation differs by $0.8\sigma$ ($1.6\,\sigma$), and the slope parameter differs by $1.7\,\sigma$ ($1.7\,\sigma$) for the \ac{sptsz} (\planck) sample. Alternatively, we fix $\by=1.67$ ($1.78$) to limit the impact of the large uncertainty on the slope on the constraint of the normalisation. In this case, we find $\ay=1.33^{+0.34}_{-0.31}$ ($1.37^{+0.36}_{-0.32}$) and the discrepancy on $\ay$ is $1.5\,\sigma$ ($3.1\,\sigma$) for the \ac{sptsz} (\planck) sample.  As in the $\zeta$-mass relation, there is no strong statistical evidence that the \ac{sptsz} clusters at low mass behave differently than those at high mass.  Tighter constraints on the high mass \ac{sptsz} scaling relation will be helpful to understand the tension.  

\begin{figure}
\hskip-0.15in\includegraphics[width=3.5in]{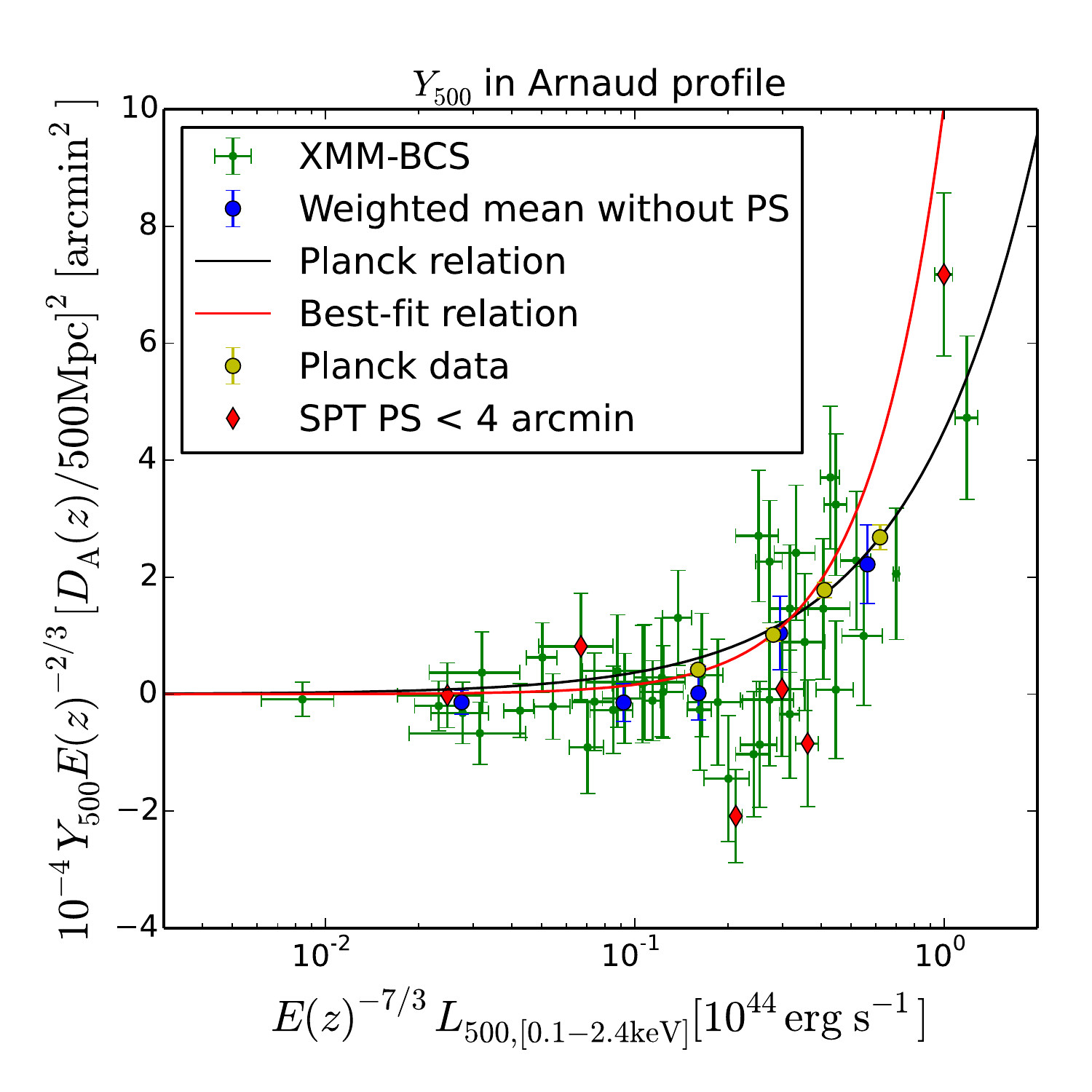}
\caption[Comparison with the \planck\ $\ysz-\lx$ relation]{Comparison with the \planck\ $\ysz-\lx$ relation.  The green dots are \xbcs\ clusters with $1\,\sigma$ uncertainty on $\SN$ and measured uncertainties on $\lx$ converted from the 0.5--2~keV band.  Blue points are inverse variance weighted means of ensembles of the \xbcs\ sample.  The black line is the \planck\ \ac{sze} relation from table 4 in \plx\ with the last four binned data points from figure~4 (\plx) in cyan.  The red line is the best-fit relation from the SPT sample.  The correction of selection bias leads to a higher-than-measured $\ysz$\ at high mass (luminous) end as the mass function is steep.)  Consistent with our parameter constraints in \Fref{fig:xbcs-ym}, our measurements prefer a lower value than the \planck\ relation.  Clusters close to \ac{spt} point sources are marked with red diamonds.}
\label{fig:xbcs-cmpplanck}
\end{figure}

The tension with the \planck\ sample is intriguing; here we discuss several possible issues that could contribute.  One difference is in the mass ranges probed in the two studies.  In \plx, the \planck\ team studies the relation between X-ray and \ac{sze} properties of 1600 clusters from the Meta-Catalogue of X-ray detected Clusters of galaxies \cite[{MCXC},][]{piffaretti11} that span two decades in luminosity ($10^{43}~\mathrm{erg\ s^{-1}} \lesssim L_{500,[0.1\ -\  2.4~\mathrm{keV}]} E(z)^{-7/3}\lesssim 2\times10^{45}~\mathrm{erg\  s^{-1}}$).  In contrast, our sample spans the range $10^{42}\mathrm{erg\  s^{-1}} \lesssim L_{500,[0.1\ -\ 2.4~\mathrm{keV}]} E(z)^{-7/3}\lesssim 10^{44}~\mathrm{erg\ s^{-1}}$ extending into the galaxy group regime.  Thus, it is interesting to probe for any mass trends in the discrepancy.  In \Fref{fig:xbcs-cmpplanck}, we show our measurements along with the \planck\ relation with fixed slope and redshift evolution as listed in Table~4 in \plx\ (solid black line).  At the luminous (massive) end, our sample matches well with the \planck\ result (cyan points are taken from Figure~4 in \plx).  Beyond the \planck\ sample at the faint end, we find the preference for lower $\ysz$ relative to the \planck\ relation.

In the \planck\ analysis, an $\lx$-mass relation without Malmquist bias correction is used  \citep{pratt09}.  They argue that based on the  similarity between the {REXCESS} and {MCXC} samples, there is no bias correction needed.  In our analysis, we use the Malmquist bias-corrected relation and our likelihood corrects for selection bias.  Using the non-corrected relation \citep{pratt09} has very little impact.  Interestingly, if we adopt the \citet{mantz10b} relation, the tension between our result and the \planck\ result disappears mainly due to the lower masses predicted by the relation as discussed in \Fref{sec:xbcs-xim}.  However, given that the \planck\ analysis adopted the \citet{pratt09} relation, it is with this same relation that the most meaningful comparisons can be made.

\begin{table}
  \caption{Constraints on the $\ysz$-mass relation.}
  \label{tab:xbcs-yszm}
  \centering
  \begin{tabular}{|crr|}
    \hline
    Parameter & $\ay[10^{-4} \mathrm{arcmin^{2}]}$&  $\by$ \\ 
    \hline
    SPT-NPS     & $1.59^{+0.63}_{-0.48}$ & $2.94^{+0.77}_{-0.74}$ \\ 
    SPT-No-SUMSS & $1.72^{+1.01}_{-0.66}$& $3.29^{+0.84}_{-0.96}$\\
    SPT           & $2.19\pm0.63$ & $1.67\pm0.29$ \\
    \planck       & $2.57\pm0.11$ & $1.78\pm0.05$ \\ 
    \hline
  \end{tabular}
\end{table}

Second, the \planck\ relation is dominated by the high mass clusters, and their measurements at the low luminosity end  (marked by cyan points in \Fref{fig:xbcs-cmpplanck}) also tend to fall below their best fit relation.  The lowest luminosity \planck\ point has a $\ysz$\ that is 68~per cent  (2\,$\sigma$ offset) of the value of the best fit model at the same X-ray luminosity.  Interestingly, the best fit normalisation of the SPT-NPS sample is 53~per cent of the \planck\ model normalisation. In this sense, the tension between the two low mass samples is less than the tension between our sample and the best-fitting \planck\ relation.

Third, we note the redshift dependence of $\ysz$-mass relation could lead to a different normalisation because the SPT-XBCS sample is on average at higher redshift than the \planck\ sample.  In \plx, they show a weak redshift evolution of $\ysz$, where the index of $E(z)$ term is $-0.007\pm 0.518$.  When they fit with the redshift evolution fixed to the self-similar expectation ($2/3$), it changes the $\ysz$ normalisation by $-5$~per cent ($0.451/0.476$), because $E(z)$ is larger than 1 for $z>0$.  In comparison, if we assume an index of 0 for $E(z)$ it will increase our $\ysz$ normalisation by 19~per cent compared to the $E(z)^{2/3}$ case (\ac{xbcs} sample has a mean redshift of $0.48$).  In this sense, there is some systematic uncertainty in the tension between the two samples that depends on the true redshift evolution of the $\ysz$-mass relation.  If the samples evolve self-similarly, then the \planck\ normalisation should be reduced by 5~per cent.

Finally, the comparison to \planck\ is complicated because of differences between the \ac{spt} and \planck\ instruments and datasets and also differences between the analyses.  Our analysis of \ac{sptsz} data calculates the \ac{sze} signal exclusively at frequencies below the \ac{sze} null (95 GHz and 150 GHz), where the \ac{sze} signal is negative, while \planck\ also includes information from frequencies above the 220 GHz \ac{sze} null, where the signal is positive.  Thus, contamination from sources like radio galaxies with steeply falling spectra, which primarily affect the lowest-frequency bands in both instruments, would tend to bias both the \planck\ and \ac{sptsz} relations in the same way.  But there are other possible sources of contamination such as dusty star-forming galaxies that are much brighter at higher frequencies.  A population of star-forming galaxies associated with clusters could artificially increase the \planck\ measured $\ysz$, but could only negatively bias the \ac{sptsz} measurements.  In their paper, the Planck team shows that at the low mass end (stellar mass smaller than $10^{11.25}\msun$), the $\ysz$ estimated by six high frequency bands directly is higher than the $\ysz$ estimated when using a thermal dust model and the $\ysz$ estimated just using the three low frequencies (100, 143, 217~GHz) \citep{planck13-11}.  However, at higher masses where we are seeing the discrepany between the SPT and \planck\ signals there is no clear evidence for a dust related systematic in the cross-checks carried out by the \planck\ team.  We present 2.8$\sigma$ significant evidence for dusty galaxy flux in our cluster ensemble in \Fref{sec:xbcs-PScontamination} below.   If present, this flux is likely contributing to some degree to the discrepancy we find between the SPT and \planck\ SZE signatures on these mass scales.

In summary, there are several potential contributing factors to the 2.8$\sigma$ tension between the two results.  None of them provide a convincing explanation for the offset on their own, but there are indications that differential sensitivity to dusty galaxy flux in SPT and \planck\ could be playing a role.  What is needed next is a larger sample with higher quality data to probe this tension and -- if the tension persists -- to provide insights into the underlying causes of the discrepancy.

\subsection{Potential Systematics}\label{sec:xbcs-sensitivity}

In the likelihood approach, we fix the cosmological parameters and assume no redshift uncertainty to improve the efficiency of the calculation.  We test both of these assumptions and find that neither significantly impacts the analysis.  Specifically, the mass function used for correcting the sampling bias is adopted from a fixed cosmology $(\Omega_{\mathrm M}, \Omega_{\Lambda},H_{0})=(0.3,0.7,70~\text{km s}^{-1}\text{Mpc}^{-1})$.  When we alter these to the recent WMAP results for $\Lambda$CDM \cite[]{komatsu11}, we find a negligible impact.

We test the importance of possible photometric redshift biases by shifting the redshifts of all clusters up (or down) by  $1\,\sigma$.  We update $\lx$ appropriately for the new redshifts, and we find a small ($0.5\,\sigma$) shift in the normalisation and no change to the slope. Therefore, redshift biases at this level would not significantly bias the analysis.

\subsection{Point Source Population}\label{sec:xbcs-PScontamination}

As already noted (see Section~\ref{sec:xbcs-xim}), there is a tendency for the systems with the most negative $\SN$ to be those with nearby \ac{spt} point sources (see \Fref{fig:xbcs-corrbias}).  In this section, we explore this association in more detail, testing whether it is biasing our constraints on the \ac{sze} mass--observable relations.  For the purposes of our analysis, an object is identified as an \ac{spt} point source if it appears as a 4$\,\sigma$ detection in a single frequency point-source filtered \ac{sptsz} map in any of the three bands (95, 150, or 220~GHz).  An area within a 4~arcmin radius around each point source is defined, and all X-ray selected clusters within that region are flagged.  There are six clusters flagged in our sample, and these are denoted with red diamonds in the figures presented above.  Given the number densities of the \ac{spt} point sources (6~deg$^{-2}$ in this field) and the X-ray selected clusters together with the association radius, we estimate a 36~per cent chance that these point sources are random associations with the clusters.

If we consider a smaller 2~arcmin association radius between the X-ray centre and the \ac{spt} point source location, we still find four associations: three of which correspond to the most negative $\SN$ in \Fref{fig:xbcs-corrbias}, and the fourth is detected only at 220~GHz by \ac{spt} (and therefore is likely a dusty galaxy).  With the smaller association radius the probability of a random association drops to 7~per cent, providing $\sim2\,\sigma$ evidence that these point sources are physically associated with the X-ray selected groups.

To further study the point source issue, we cross-match our cluster sample with radio sources detected at 843~MHz by the \ac{sumss}. The survey covers the whole sky at $\delta\leq-30^{\circ}$ with $|b|>10^{\circ}$ down to limiting source brightness of 6~mJy~beam$^{-1}$.  For the cross-matching, we utilise the latest version 2.1 of the catalog\footnote{\url{http://www.physics.usyd.edu.au/sifa/Main/SUMSS}} and a similar matching radius of 2~arcmin.  This threshold is much larger than the \ac{sumss} positional uncertainty, which has a median value of $\sim2.3$~arcsec.

Within 2~arcmin of the X-ray centres, we find a total of 19 \ac{sumss} point sources matching 18 clusters from our sample.  In comparison, given the number density of \ac{sumss} sources \cite[$31.6$~deg$^{-2}$,][]{mauch03}, the number density of our clusters, and our association radius, we would expect to find $\sim 5$ clusters randomly overlapping with point sources in the $6~\sdeg$ survey; there is a $3\times10^{-4}$~per cent chance of explaining the associations as random superpositions.
 Thus, our small sample provides clear evidence of physical associations between low frequency radio point sources and X-ray selected groups and clusters; this is consistent with previous findings \cite[]{best05,lin07} that low frequency radio sources are associated with cluster galaxies in both optically and X-ray selected cluster samples.  As expected, given the tendency for radio galaxies to have steeply falling spectra as a function of frequency, only a small fraction (3 out of 19) of these low frequency radio galaxies are detectable at \ac{spt} frequencies.

We use the \ac{bcs} data \citep{desai12} to examine the optical counterparts of the six \ac{spt} point sources that lie within 4~arcmin of our X-ray selected group and cluster sample.  We do this by first associating the \ac{spt} point sources with a \ac{sumss} source, which in general is only possible for the radio galaxies and not the dusty galaxies \citep{vieira10}.  For our sample, three of the \ac{spt} point sources within 4~arcmin of the X-ray selected groups and clusters  have \ac{sumss} counterparts.   All three of these have strongly negative $\SN$ (see \Fref{fig:xbcs-corrbias}).  For two of the three point sources, the optical counterpart is the group \acs{bcg}.  In the third case the \ac{spt} point source corresponds to a quasar candidate \cite[MRC~2319-550;][]{wright90} and does not appear to be a cluster member.  The three remaining \ac{spt} point sources do not have \ac{sumss} counterparts and are likely dusty galaxies;  the \ac{sze} signatures $\SN$ of those systems are not obviously impacted.  Thus we confirm that in two of our 46 low mass systems there are associated radio galaxies bright enough to be detected at \ac{spt} frequencies.

Based on the prediction from \cite{lin09}, we would have expected that radio sources completely fill in the $Y_\mathrm{SZ}$ signal (100~per cent contamination) at a redshift of 0.1 (or a redshift of 0.6) in approximately 2.5 (or 0.5) percent of clusters with similar mass ($M_{200}=10^{14}\msun$).  For our 46 cluster sample, we would have expected this to happen for 1.15 (or 0.23) clusters, consistent with the two clusters we find associated with radio galaxies detected as point sources by \ac{sptsz}.  We also expect a 20~per cent level $Y_\text{SZ}$ contamination on 9 (2)~per cent of the sample. This predicted contamination is significantly smaller than our current uncertainties on the  $Y_\mathrm{SZ}$  normalisation, and therefore cannot be tested in this analysis.

We repeat the \ac{sze}-mass relation analysis while excluding the half of the clusters with \ac{sumss} point source associations.  We find that the results are qualitatively similar using either the SPT-NPS or SPT-No-SUMSS sample (see  Tables~\ref{tab:xbcs-constraint} and \ref{tab:xbcs-yszm}), although the uncertainties increase; this is consistent with the expectation that the level of the effect is too small to be measured with our sample.  As already shown in Tables~\ref{tab:xbcs-constraint} and \ref{tab:xbcs-yszm}, our analysis shows no statistically significant difference in the \ac{sze}-mass relations when excluding or including the systems with nearby \ac{spt} point sources.

As pointed out in \Fref{sec:xbcs-yszm}, the dusty star-forming galaxies would have a net negative biasing impact on the \ac{sptsz} measurement.  We examine the contamination from the dusty galaxies, which are not bright enough to be directly detectable in the 150~GHz and 95~GHz bands.  To do this we measure the specific intensities at 220~GHz in a single frequency adaptive filter that uses cluster profiles at the locations of our X-ray selected cluster sample.  In the SPT-NPS sample, the evidence for dusty galaxies is significant at the 2.8\,$\sigma$ level.  We then convert the 220~GHz intensities to temperature fluctuations at 150~GHz and 95~GHz by assuming the intensity follows $I\propto \nu^{3.6}$ for dusty sources \cite[]{shirokoff11}.  These are then converted to the corresponding values of $\ysz$.  Dividing then by the expected $\ysz$ for a cluster of this redshift and X-ray luminosity, we then estimate the inverse variance weighted mean contamination to be $32\pm18$~per cent and $7\pm4$~per cent at 150~GHz and 95~GHz, respectively.  Together, this contamination would lead the \ac{sptsz} observed $\ysz$ signature to be biased low by $\sim(17\pm9)$~per cent.  This fractional contamination depends on the mass and redshift of the cluster together with the typical star formation activity.  In particular, as a function of mass the SZE signature grows as $\ysz\propto M^{5/3}$, whereas the blue or star forming component of the galaxy population falls \citep[e.g.][]{weinmann06};  thus, contamination would fall with mass.  As one pushes to even higher redshift than this sample (i.e. $z>1$) where star formation is more prevalent, the contamination would be expected to increase.

This level of contamination is consistent with a recent study of $\sim 550$ galaxy clusters selected via optical red-sequence techniques.  Using \emph{Herschel} and \ac{spt} mm-wave data to jointly fit an \ac{sze}+dust spectral model, \cite{bleem13} finds the contamination at 150~GHz  to be $40 \pm 30$~per cent for low-richness optical groups ($M_{\textrm{200}} \sim 1 \times 10^{14}\msun$). The fractional contamination declines as a function of optical richness and is measured to be $5 \pm 5$~per cent for the richest 3 per cent of clusters in the sample sample ($M_{\textrm{200}} \sim 3 \text{--} 6 \times 10^{14}\msun$).  A larger sample size combined with deeper mm-wave data will improve our ability to estimate the contamination from dusty galaxies in clusters and groups.

In summary, this small sample of 46 X-ray selected groups and low mass clusters provides high significance evidence of having physically associated low frequency SUMSS radio galaxies.  For the \ac{spt} point source sample within 2~arcmin, there is less than $2\,\sigma$ statistical evidence of physical association, but two of the sources have optical counterparts that are in the groups.  Although we would expect physically associated high frequency radio  galaxies to bias the \ac{sze} mass-observable relation, our analysis provides no evidence of this impact.  We use the 220~GHz \ac{sptsz} data in this sample to estimate that the $\ysz$ measured by the \ac{spt} is biased $\sim 17\pm 9$~per cent low.  A larger sample from a broader survey (through XMM-XXL or eROSITA, for example) or a deeper \ac{sze} survey would both help to improve our understanding of the impact of point sources.

\section{Conclusions}\label{sec:xbcs-conclusions}

Using data from the \ac{sptsz} survey, we have explored the \ac{sze} signatures of low mass clusters and groups selected from a uniform {\it XMM-Newton} X-ray survey.  The cluster and group sample from the \ac{xbcs} has a well understood selection, and previously published calibrations of the $\lx$-mass relation allow us to estimate the masses of each of these systems. Although these systems have masses that are too low for them to have been individually detected within the \ac{sptsz} survey, we are able to use the ensemble to constrain the underlying relationship between the halo mass and the \ac{sze} signature for low mass systems.

Our method corrects for the Eddington bias and shows that there is no Malmquist like bias effect on the \ac{sze} mass-observable relation within this X-ray selected sample.  We test our likelihood using a large mock sample, and we show with the current sample size we can at most extract constraints from two scaling relation parameters: the power law amplitude $\asz$ and slope $\bsz$ (see Equations~\ref{eq:zetam} and \ref{eq:xbcs-ym}).

We separate the sample of 46 groups and clusters into three subsamples: (1) the full sample, (2) the point source-free sample, for which we exclude systems with point sources detected at significance $>$4 at either 95, 150, or 220~GHz in the \ac{sptsz} data within 4~arcmin radius of the X-ray centre, and (3) the point source-free sample, with clusters at $z<0.3$ excluded.  We find that, due to the point source contamination in three of the lowest $\SN$ groups, the full sample exhibits a steep slope ($\bsz=2.80^{+0.66}_{-0.63}$) that is in tension at 2.6$\,\sigma$ with the high mass \ac{spt} sample ($\bsz=1.40\pm0.16$).   The point source free subsample has a slope ($\bsz=2.14^{+0.86}_{-0.66}$) that is in rough agreement with the slope of the high mass \ac{spt} sample ($1.4\sigma$ difference).  We find no evidence that the low redshift clusters deviate from the scaling relation of the point source free sample. 

We also measure the $\ysz$-mass relation for our sample and compare it to the results from the \ac{sptsz} high mass clusters and the \planck\ sample.  Our low mass sample exhibits a preference for lower normalisation and steeper slope than the other two samples, but the uncertainties are large (see \Fref{fig:xbcs-ym} and \Fref{tab:xbcs-yszm}).  Within the \ac{spt} samples, there is no statistically significant evidence for differences in the scaling relation as one moves from high to low masses.  On the other hand, the \planck\ sample exhibits a 2.8\,$\sigma$ significant tension with our sample.  As shown in \Fref{fig:xbcs-cmpplanck}, the lowest X-ray luminosity portion of our sample has lower $\ysz$ than expected from the \planck\ relation. We discuss a range of possible explanations for this tension (Section~\ref{sec:xbcs-yszm}), in particular contamination from dusty sources.  Given the significance level of the tension the appropriate next step is to enlarge the sample to better quantify the differences in the \ac{sze} signatures of low and high mass clusters and the possible differences between \planck\ and \ac{spt}.

We examine radio point source contamination.  Cross-matching our X-ray selected groups and clusters with the \ac{sumss} catalog, we find that 18 of 46 members have associated  843~MHz \ac{sumss} point sources within 2~arcmin.  This represents highly significant evidence of physical association between our sample and low frequency point sources.  At higher frequencies, we find four systems with associated \ac{spt} detected point sources; three of these also have SUMSS counterparts.  Two of these three point sources have optical counterparts that lie within the X-ray group, and the third is a quasar candidate that is likely unassociated with the group.  Having two out of 46 groups or clusters with physically associated bright, high frequency point sources is consistent with the expectations from \citet{lin09}.  The predicted contamination from undetected radio point sources \citep[]{lin07,lin09} in the remainder of the sample is significantly smaller than our measurement uncertainty on the $\ysz$ normalisation, and so we cannot test these predictions here.  

We also examine the impact of undetected dusty galaxies.  Using the \ac{sptsz} 220~GHz band, we find 2.8$\,\sigma$ significant evidence of a flux excess due to dusty galaxies.  Extrapolating to lower frequencies, we estimate that the measured $\ysz$ signature is biased low by $\sim(17\pm9)$~per cent in this ensemble of low mass clusters and groups.  Given the different frequency coverage of \planck\ and \ac{spt}, it is not clear that the \planck\ bias due to dusty galaxy flux would be the same.   If flux from dusty galaxies would induce a smaller negative bias or even a positive bias in \planck\ $\ysz$ measurements, then that would reduce the tension between the \planck\ $\ysz$--mass relation and ours.

We point out that these contamination levels are for this X-ray selected low mass sample with a median mass of $10^{14}\msun$, which is significantly below the typical mass of \ac{spt} selected clusters.  Given the increasing rarity of blue and star forming galaxies as one moves from groups to high mass clusters \citep[e.g][]{weinmann06}, any contamination in the \ac{spt} selected sample would be much lower.
 
Finally, the receiver on the \ac{spt} was upgraded in 2012. The SPTpol camera provides sensitivity to \ac{cmb} polarization and, more importantly for \ac{sze} work, increased sensitivity to CMB temperature fluctuations. The final SPTpol maps are expected to cover 500 square degrees of sky to noise levels of $\sim 5$ and $\sim 9 \mr{\mu K-arcmin}$ at 150 and 95~GHz \citep{austermann12}.  Meanwhile, the XXL survey \cite[]{pierre11} has increased the survey area that has a characteristic 10~ks {\it XMM-Newton} exposure from $6~\sdeg$ to $25~\sdeg$.  This should enable an interesting new insight into possible differences in the \ac{sze} signatures of low and high mass clusters.  We make a forecast with a mock catalog that consists of 144 clusters within redshift range 0.2--1.2 and a bolometric flux limit of $1\times10^{-14}~\text{erg s}^{-1}\text{cm}^{-2}$.  Analysing this sample with the appropriate SPTpol increase in depth indicates that with the future sample we can tighten the fractional error on $\asz$ to 6~per cent compared to our current result of 30~per cent.  On $\bsz$ the uncertainty shrinks from 34 to 8~per cent.  These improvements should enable a more revealing comparison of the \ac{sze} signatures of low and high mass clusters and perhaps also enable a detailed study of potential contamination of the \ac{sze} signal by associated radio or dusty galaxies.

 \section*{Acknowledgments}
We acknowledge the support of the DFG through TR33 ``The Dark Universe'' and the Cluster of Excellence ``Origin and Structure of the Universe''.  Some calculations have been carried out on the computing facilities of the Computational Center for Particle and Astrophysics (C2PAP).  The South Pole Telescope is supported by the National Science Foundation through grant PLR-1248097.  Partial support is also provided by the NSF Physics Frontier Center grant PHY-1125897 to the Kavli Institute of Cosmological Physics at the University of Chicago, the Kavli Foundation and the Gordon and Betty Moore Foundation grant GBMF 947.  This work is also supported by the U.S. Department of Energy. Galaxy cluster research at Harvard is supported by NSF grants AST-1009012 and DGE-1144152. Galaxy cluster research at SAO is supported in part by NSF grants AST-1009649 and MRI-0723073. The McGill group acknowledges funding from the National Sciences and Engineering Research Council of Canada, Canada Research Chairs program, and the Canadian Institute for Advanced Research.

\bibliographystyle{mn2e} \bibliography{spt}

\appendix

\section{Likelihood function}
\label{sec:likelihood}
We start from the full likelihood function based on B13 to constrain both the cosmological model and the scaling relations as (note that the observables are different from the ones used in B13):
\begin{align}
\ln\ \mathcal{L}(\bm{c},\rsz,\rx,\Theta) &=
\sum_{i}\ln\frac{\mr{d}N(Y_{i}, f_{i}, z_{i}|\bm{c},\rsz,\rx,\Theta)}{\mr{d}Y\mr{d}f\mr{d}z}\nonumber\\ 
 &- \iiint\frac{\mr{d}N(Y, f, z|\bm{c},\rsz,\rx,\Theta)}{\mr{d}Y\mr{d}f\mr{d}z}\mr{d}Y\mr{d}f\mr{d}z,\label{eq:xbcs-full-likelihood}
\end{align}
where $i$ runs over the cluster sample, $Y_i$ is the \ac{sze} signal (i.e. $\SN$ or $\ysz$), $f_i$ is the X-ray flux, and $z_i$ is the redshift.  $\rsz$ represents the \ac{sze} scaling relation, $\rx$ represents the X-ray scaling relation, and $\Theta$ describes the sample selection.  $\mr{d}N(Y_{i}, f_{i}, z_{i}|\bm{c},\rsz,\rx,\Theta)$ is the expected number of clusters within a three-dimensional cell $\mr{d}Y\mr{d}f\mr{d}z$, and the second term is the integral of the differential cluster number density over all $Y$, $f$ and $z$.

Given the limited sample size, we focus on the \ac{sze}-mass scaling relation, keeping the cosmological $\bm{c}$ and the X-ray scaling relation $\rx$ fixed.  In addition, we assume the redshift measurements have insignificant uncertainties.  Within this context, the X-ray flux is equivalent to the X-ray luminosity $L$.   

The differential number density of clusters can be expressed as:
\begin{align}
&\frac{\mr{d}N(Y, L, z|\bm{c},\rsz,\rx,\Theta)}{\mr{d}Y\mr{d}L\mr{d}z}\nonumber \\
&\ \ \ \ \ = P(Y|L,z,\bm{c},\rsz,\rx, \Theta)\,\frac{\mr{d}N(L, z| \bm{c}, \rsz, \rx, \Theta)}{\mr{d}L\mr{d}z}, \label{eq:xbcs-conditional-yl}
\end{align}
where the first factor is the conditional probability of $Y$ given observables $L$ and $z$ with other model parameters, and we are using the relation $\mr{d}N/\mr{d}Y=P(Y)N$.  The second factor is the differential number density of clusters as a function of $L$ and $z$.

The full likelihood can be split into three parts:
\begin{align}
\ln \mathcal{L}(\bm{c},\rsz,\rx,\Theta)=& \sum_{i}\ln P(Y_{i}|L_{i},z_i,\bm{c} ,\rsz,\rx, \Theta) \nonumber \\
+& \sum_{i}\ln \frac{\mr{d}N(L_{i}, z_{i}|\bm{c} ,\rsz,\rx, \Theta)}{\mr{d}L\mr{d}z} \nonumber\\ 
-& \iiint\frac{\mr{d}N(Y, L, z|\bm{c} ,\rsz,\rx, \Theta)}{\mr{d}Y\mr{d}L\mr{d}z}\mr{d}Y\mr{d}L\mr{d}z. \label{eq:full-likelihood-3}
 \end{align}
If the sample selection is based on the X-ray only, then we have:
\begin{equation}\label{eq:xbcs-selection-x}
\mr{d}N\left(L_{i},z_{i}|\bm{c},\rsz,\rx,\tx\right)=\tx(L_{i},z_{i})\mr{d}N(L_{i},z_{i}|\bm{c},\rx),
\end{equation}
where $\tx$ is simply the probability that a cluster with X-ray luminosity $L_i$ and redshift $z_i$ is observed.  In addition,
\begin{equation}\label{eq:xbcs-normalise-yl}
\int P(Y|L,z,\bm{c} ,\rsz,\rx, \tx) \mr{d}Y = 1,
\end{equation}
which simply means that, because there is only X-ray selection $\tx$, any cluster that makes it into the sample due to its X-ray properties will always have a corresponding value $Y$.   Using this condition together with \Fref{eq:xbcs-conditional-yl} allows us to write the third term in \Fref{eq:full-likelihood-3} as: 
\begin{align}\label{eq:xbcs-full-liklihood-3rd}
&\iiint\frac{\mr{d}N(Y, L, z|\bm{c} ,\rsz,\rx, \tx)}{\mr{d}Y\mr{d}L\mr{d}z}\mr{d}Y\mr{d}L\mr{d}z\nonumber\\
&=\iint\frac{\mr{d}N(L, z|\bm{c} ,\rx, \tx)}{\mr{d}L\mr{d}z}\mr{d}L\mr{d}z.
\end{align}
Note that by adopting Equations~(\ref{eq:xbcs-selection-x}) and (\ref{eq:xbcs-full-liklihood-3rd}), the last two terms in \Fref{eq:full-likelihood-3} have no remaining dependence on $Y$ and depend only on cosmology $\bm{c}$, the X-ray-mass scaling relation $\rx$ and the X-ray sensitive selection $\tx$.  Thus, within the context of a fixed cosmology and X-ray scaling relation these two terms are constant and do not contribute to constraining the \ac{sze} scaling relation $\rsz$.  Thus, for the final likelihood that we use in this analysis, we obtain
\begin{equation}\label{eq:xbcs-final-likelihood}
\ln \mathcal{L}(\rsz)
= \sum_{i}\ln P(Y_{i}|L_{i},z_{i},\bm{c},\rx,\rsz,\tx).
\end{equation}
The derivation of the likelihood is correct even in the presence of correlated scatter between $L$ and $Y$.

However if the selection were based on both $L$ and $Y$, then \Fref{eq:xbcs-final-likelihood} would no longer be equivalent to the full likelihood.  For instance \Fref{eq:xbcs-selection-x} would need to be extended as:
\begin{equation}\label{eq:xbcs-selection-yl}
\mr{d}N\left(L_{i},z_{i}|\bm{c},\rsz,\rx,\Theta\right)=\int \mr{d}Y \Theta(Y,L_{i},z_{i}) \mr{d}N(Y, L_{i},z_{i}|\bm{c},\rx, \rsz).
\end{equation}
And therefore detailed modelling of the selection would be required to calculate the likelihood and constrain the scaling relation parameters.  

\end{document}